\pdfoutput=1

\documentclass[aps, prb, reprint, showpacs]{revtex4-1}

\usepackage{amsmath, amssymb}
\usepackage{times, mathptmx}
\usepackage{graphicx}
\usepackage{hyperref}

\newcommand{\kk}{\textbf{k}}

\renewcommand{\Im}{\mathop{\mathrm{Im}}}
\renewcommand{\Re}{\mathop{\mathrm{Re}}}

\begin{document}

\date{\today}
\title{Blowup dynamics of coherently driven polariton condensates}
\author{S.~S.~Gavrilov}
\affiliation{Institute of Solid State Physics,
RAS, Chernogolovka, 142432, Russia}

\begin{abstract}
  Basing on the Gross-Pitaevskii equations, it is predicted that a
  repulsive (defocusing) interaction makes a 2D polariton condensate
  able to accumulate its energy under above-resonance optical pumping.
  The energy can be accumulated during a lot of polariton lifetimes,
  resulting in the state in which the mismatch of the pump frequency
  is compensated by the blueshift of the polariton resonance.  The
  process begins when the field density reaches the parametric
  scattering threshold that is inversely proportional to the polariton
  lifetime.  Although the increase in energy may be arbitrarily slow
  in its beginning, it is followed by a blowup.  This scenario applies
  to the case of the transitions between steady states in multistable
  cavity-polariton systems.  There is a tradeoff between the latency
  of the transitions and the pump power involving them.
\end{abstract}


\maketitle

\section{Introduction}
\label{sec:intro}

This study is devoted to the problem of non-equilibrium transitions in
multistable cavity-polariton systems.  Bi- and multistability of
nonlinear media in which gain or decay rate ($\gamma$) of light shows
a threshold behavior had been studied for decades.  Less known is
another type of bistability that was predicted\cite{Elesin73} to occur
under resonant optical excitation of a macroscopically coherent state
of bosons with a spin of 1 and a repulsive two-particle interaction,
such as exciton gas in semiconductors.  Due to the interaction the
condensate level~$\hbar \omega_c$ shows a blueshift with increasing
its population density~$n$.  If the frequency of the pump wave
exceeds~$\omega_c$, the positive feedback loop between $n$ and
$\omega_c$ makes the system unstable within a finite interval of $n$.
This kind of optical bistability, stemming from ``intrinsic''
interactions in a Bose gas rather than a non-linearity of a medium,
has recently been observed in cavity-polariton
systems.\cite{Baas04-pra, Gippius04-epl, Gippius04-cm, Paraiso10,
  Sarkar10, Adrados10, Gavrilov10-jetpl-en, Gavrilov12-prb}

Cavity polaritons are composite bosons trapped in a cavity active
layer due to the strong exciton-photon
coupling.\cite{Weisbuch92,Yamamoto-book-2000,Kavokin-book-03} Their
lifetime is very small ($\tau \sim 10^{-12}$\,--\,$10^{-11}$~s in GaAs
based microcavities); yet their interaction strength provides a
blueshift that exceeds the resonance width ($\hbar \gamma = \hbar /
\tau$) even at comparatively low pump densities where the system can
still be considered as a weakly non-ideal gas of bosons.  The
steady-state intensity of the driven mode can vary over more than an
order of magnitude near critical (threshold) values of the pump
density.\cite{Paraiso10,Sarkar10} Polariton multistability allows
switching the cavity between distinct steady states those, in a
general case, differ in both intensity and
polarization.\cite{Gippius07, Shelykh08-prl, Gavrilov10-en,
  Gavrilov10-jetpl-en} Recently there have been reported the
transitions between linear and circular as well as right- and
left-circular polarization states, which proceed on the scale of
picoseconds under a constantly polarized pump
wave.\cite{Gavrilov13-apl,Gavrilov13-mf,Sekretenko13-fluct} The
minimum switching time is comparable to $\tau$\cite{Sekretenko13-10ps}
and the minimum size of a ``multistable cell'' is several
microns.\cite{Paraiso10} In its turn, the multistability gives rise to
a spectacular row of collective phenomena in polariton physics, such
as self-organized optical parametric oscillation
(OPO),\cite{Demenev08,Krizhanovskii08} spin
rings,\cite{Shelykh08-prl,Sarkar10,Adrados10} threshold-like screening
of surface acoustic waves,\cite{Krizhanovskii13} and bright polariton
solitons.\cite{Egorov09,Egorov10-prl,Sich13} The combination of
all-optical tunability, compactness and a high speed of transitions
makes cavity-polariton systems interesting for applications in the
field of digital processing.

In this work we explore the dynamics of the bistability-initiated
transitions.  Using the Gross-Pitaevskii equations, we have found that
a passage from the low- to high-energy state is mediated by the
parametric scattering into signal/idler modes whose in-plane wave
vectors differ from that of the pumped mode.  Such scattering, which
also shows a threshold onset with increasing pump power and is known
to result in OPO states under pumping near the ``magic
angle'',\cite{Stevenson00, Ciuti01, Whittaker01, Butte03, Whittaker05}
was not, however, taken into account in the studies of the
multistability under normal-incidence pump.\cite{Baas04-pra,
  Gippius07, Shelykh08-prl, Gavrilov10-en} Here we show that
inter-mode scattering can have a drastic impact upon the
multistability thresholds and duration of the transitions between
steady states.  The scattered modes (``signals'') are fed by the
pumped mode that starts to break-up above the scattering threshold;
importantly, this process is accompanied by a growth of both the
``signal'' and pumped mode amplitudes even at constant pump power.
Although the growth can be arbitrarily slow in its beginning, with
time it is followed by an explosive amplification of the pumped mode
and a transition to the upper branch of stability.  This effect is
essentially collective; it cannot be reproduced within the framework
of three-mode OPO models with fixed ``signal'' and ``idler'' wave
vectors.

Similar scenarios, which imply a hyperbolic growth, or a singularity
being reached in a finite time, are often referred to as regimes (or
solutions) \emph{with blowup}.  Besides various models in many areas
of knowledge, they are predicted to exist in systems described by the
Schr\"odinger equation with cubic nonlinearity, also known as the
Gross-Pitaevskii equation.  If the system is conservative and the sign
of the nonlinear term corresponds to \emph{attractive} interaction,
then singular solutions occur due to an intense self-focusing of the
field.\cite{McLaughlin86,Landman88} By contrast, in this work we face
a qualitatively new type of blowup behavior that takes place in
systems described by the Gross-Pitaevskii equations with (i) a
\emph{repulsive} interaction and (ii) allowance made for both
dissipation and coherent driving.

The paper is organized as follows.  In Sec.~\ref{sec:thresholds} we
analyze the effects of bistability and parametric scattering.  The
method is based on the Bogoliubov approximation and is widely used for
studying OPO states in cavity-polariton systems.\cite{Ciuti01,
  Whittaker01, Gippius04-epl, Gippius04-cm, Carusotto04, Whittaker05,
  Wouters07-thr, Gavrilov07-en} Here our particular aim is to
establish the relation between the scattering and bistability
thresholds for the case of pumping near normal incidence.  In
Sec.~\ref{sec:catastrophe} we study the condensate above the scattering
threshold and prove that no steady states can be formed below the
catastrophe point in which intra-cavity field grows explosively; that
is the central point of our work.  Section~\ref{sec:main-results},
which can be read independently of Sec.~\ref{sec:catastrophe}, contains
a qualitative description and discussion of the main results. In
Sec.~\ref{sec:num} we give a numerical example that illustrates the
considered evolution scenario.  Finally, Sec.~\ref{sec:conclusion}
contains concluding remarks.

\section{Bistability vs.\ parametric scattering: setting up the
  problem}
\label{sec:thresholds}

The dispersion law for cavity polaritons has the form
\begin{multline}
  \label{eq:dispersion}
  \omega_\mathrm{LP,UP}(\kk) = \frac12 [\omega_C(\kk) + \omega_X(\kk)]
  \\ {} \mp \frac12 \sqrt{[\omega_C(\kk) - \omega_X(\kk)]^2 + 4g^2},
\end{multline}
where $\omega_{C,X}^{\mathstrut}(\mathbf{k}) = \omega_{C,X}^{(0)} +
\hbar \mathbf{k}^2 / 2 m_{C,X}^{\mathstrut}$ are the 2D cavity-photon
$(C)$ and exciton $(X)$ frequencies, $\mathbf{k}$ the in-plane wave
vector, $g$ the exciton-photon coupling constant; LP and UP stand for
the lower and upper polariton branches.  The photon mass $m_C \sim
10^{-5} m_e$ is much smaller than the exciton mass $m_X \sim 10^{-1}
m_e$; therefore $\omega_X$ can be considered constant at small~$k$.
In GaAs cavities $\hbar g$ is of the order of several meV and largely
exceeds the width of both photon and exciton levels ($\hbar \gamma
\sim 0.1$~meV).  A plane wave with frequency~$\omega_p$ and incidence
angle~$\theta$ excites polaritons with~$k_p = \omega_p \sin (\theta) /
c$.  The presence of the upper dispersion branch will be further
neglected on the assumption that $|\omega_p - \omega_\mathrm{LP}(k_p)|
\ll \omega_\mathrm{UP}(k_p) - \omega_\mathrm{LP}(k_p)$.

Throughout this work we assume that the state of the polariton system
is macroscopically coherent under the conditions of coherent
(plane-wave) pumping.  Therefore its evolution is described by the
Gross-Pitaevskii equation.  In the $k$-space representation it
reads\cite{Gippius04-epl}
\begin{multline}
  \label{eq:gp}
  i \frac{\partial}{\partial t} \psi(\kk, t) = [\omega(\kk) - i
  \gamma(\kk)] \psi(\kk, t) + \delta(\kk, \kk_p) f e^{-i \omega_p t}
  \\ {} + V \sum_{\mathbf{q}_1, \mathbf{q}_2} \psi^* (\mathbf{q}_1 \,
  {+} \, \mathbf{q}_2 \, {-} \, \kk, t) \, \psi(\mathbf{q}_1, t) \,
  \psi(\mathbf{q}_2, t).
\end{multline}
Here $f$ is the pump amplitude, $\psi$ the intra-cavity field
(``macroscopic wavefunction''), $\omega = \omega_\mathrm{LP}$ the
eigenfrequency and $\gamma$ the decay rate; $V > 0$ is the strength of
the polariton-polariton interaction per unit area, $\delta$ is
Kronecker delta.  Amplitudes $f$ and $\psi$ are expressed in arbitrary
units, however, $V |\psi|^2$ has the dimension of frequency and
determines the resonance blueshift.  The spin degrees of freedom are
neglected, which corresponds to the case of pumping with circularly
polarized light,\cite{Gippius07, Gavrilov10-en} since the interaction
between opposite-spin polaritons as well as TE/TM splitting can be
considered negligible in isotropic cavities near $\kk_p = 0$.

\begin{figure}
  \centering
  \includegraphics[width=1\linewidth]{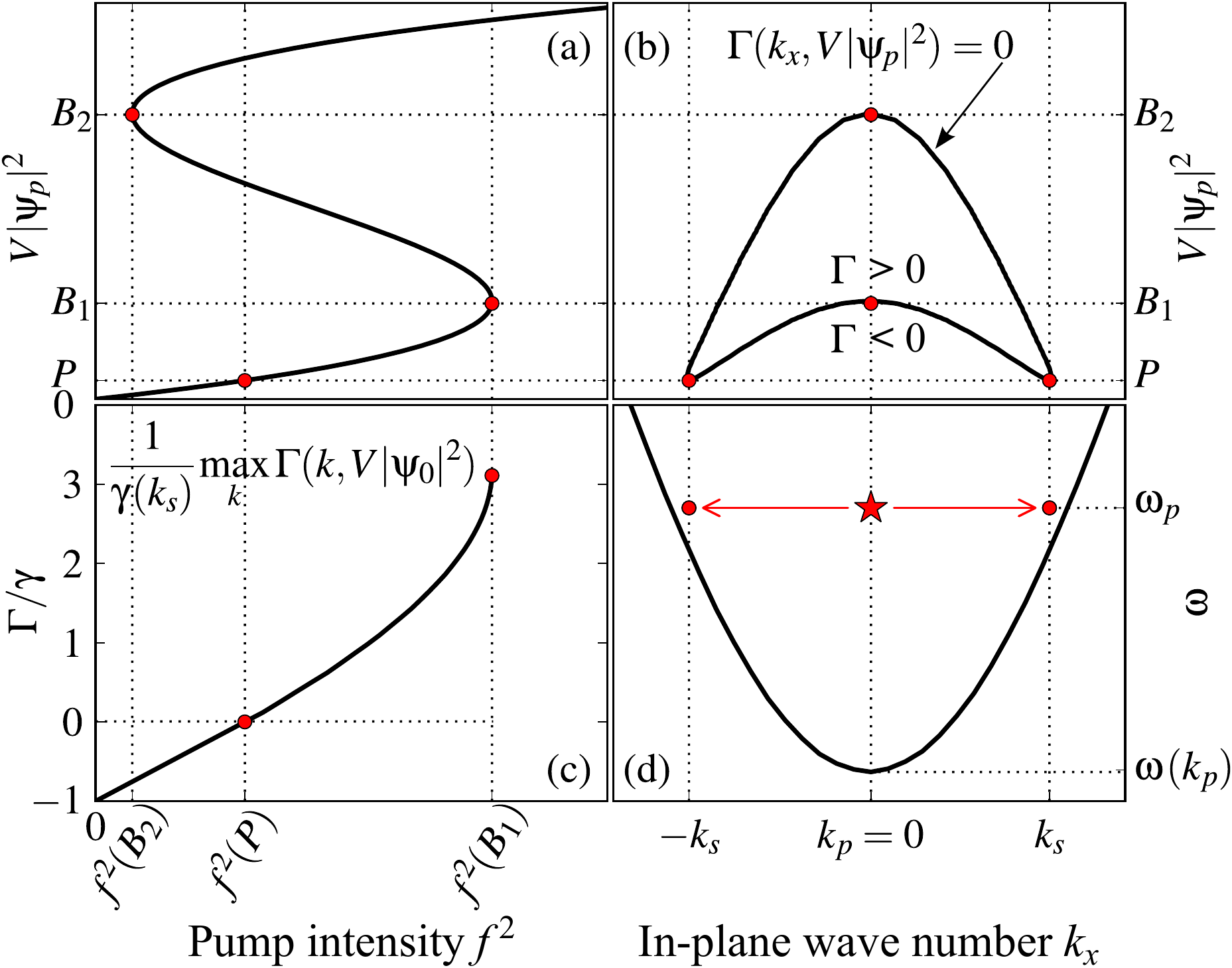}
  \caption{(a) Intra-cavity field $|\psi_p|^2$ as function of $f^2$ in
    the one-mode approximation [see Eq.~(\ref{eq:one-mode})];
    (b)~solutions (\ref{eq:sc-thr}) of equation $\Gamma(k, V
    |\psi_p|^2) = 0$ defining the boundary between parametrically
    stable ($\Gamma < 0$) and unstable ($\Gamma > 0$) modes; (c)~gain
    rate of scattered modes $\max_k \Gamma(k, V |\psi_p|^2) /
    \gamma(k_s)$ at $V |\psi_p|^2 < B_1$, expressed in units of the
    polariton decay rate; (d)~bare dispersion law and a scheme of
    parametric scattering.  The solutions are obtained at
    $\hbar[\omega_p - \omega(k_p)] = 0.5$~meV and $\hbar\gamma \approx
    0.04$~meV.}
  \label{fig:steadystate}
\end{figure}

Within the one-mode approximation, i.\,e.\ $\psi(\kk, t) = \delta(\kk,
\kk_p) \, \psi_p e^{-i \omega_p t}$, the response of the driven mode is
defined by the equation
\begin{equation}
  \label{eq:one-mode}
  |\psi_p|^2 = \frac{f^2}{\left[ \omega_p - \omega(\kk_p) - V
      |\psi_p|^2 \right]^2 + \left[ \gamma(\kk_p) \right]^2}.
\end{equation}
If $\omega_p - \omega(\kk_p) > \sqrt{3} \gamma(\kk_p)$, the dependence
of $|\psi_p|^2$ on $f^2$ has the form of an S-shaped curve
[Fig.~\ref{fig:steadystate}(a)].\cite{Elesin73,Baas04-pra,
  Gippius04-epl} The solutions with negative derivative $d(f^2) /
d(|\psi_p|^2)$ are asymptotically unstable (see below).  They lie
within the interval $B_1 < V |\psi_p|^2 < B_2$, where
\begin{equation}
  \label{eq:bs-thr}
  B_{1,2} = \frac23 [\omega_p - \omega(\kk_p)] \mp
  \frac13 \sqrt{[\omega_p - \omega(\kk_p)]^2 - 3 [\gamma(\kk_p)]^2}.
\end{equation}
Attainment of $V |\psi_p|^2 = B_1$ yields transition to the upper
steady-state branch.

Besides the bistability effect, increasing density can involve loss of
stability due to elastic two-particle scattering from $\kk = \kk_p$ to
the ``signal'' ($\kk_s, \tilde\omega_s$) and ``idler'' ($\kk_i,
\tilde\omega_i$)
modes.\cite{Ciuti01, Whittaker01, Gippius04-epl} According
to the conservation laws,
\begin{equation}
  \label{eq:cons1}
  2 \kk_p = \kk_s + \kk_i \quad\text{and}\quad
  2 \hbar \omega_p =
  \hbar \tilde\omega_s + \hbar \tilde\omega_i,
\end{equation}
or
\begin{equation}
  \label{eq:cons2}
  2 \omega_p = \tilde\omega(\kk) + \tilde\omega(2\kk_p - \kk),
\end{equation}
where $\tilde\omega = \tilde\omega(\kk)$ is a renormalized
eigenfrequency.  The scattering directions and threshold can be found
using the Bogoliubov approximation, that is
\begin{gather}
  \label{eq:bogoliubov1}
  \psi(\kk, t) = \delta(\kk, \kk_p) \, \psi_p e^{-i\omega_p t}
  + \tilde\psi (\kk) \, e^{-i\tilde\omega (\kk) t}, \\
  \label{eq:bogoliubov2}
  |\tilde\psi(\kk)| \ll |\psi_p| \quad \text{for each $\kk$}.
\end{gather}
Under the above assumptions~(\ref{eq:cons2})--(\ref{eq:bogoliubov2}),
the Gross-Pitaevskii equation~(\ref{eq:gp}) and its complex conjugate
form a $2 \times 2$ linear problem connecting $\tilde\psi(\kk)$ and
$\tilde\psi^*(2\kk_p-\kk)$.  Then solving the characteristic equation
yields frequencies $\Omega = \Omega(\kk, V |\psi_p|^2)$ of
above-condensate states.  Stationary solutions $\psi_p$ at which
$\Gamma = \Im \Omega(\kk, V |\psi_p|^2)$ takes positive values at any
$\kk$ are asymptotically unstable. Equation $\Gamma(\kk, V |\psi_p|^2)
= 0$ implicitly defines the scattering threshold $V |\psi_p|^2 = P$ as
function of the corresponding ``signal'' wave vector~$\kk$.

Figure~\ref{fig:steadystate}(b) presents a typical solution of the
equation $\Gamma = 0$ at $\kk_p = 0$; Fig.~\ref{fig:steadystate}(c)
shows the largest (over $\kk$) gain rate $\Gamma$ at $V |\psi_p|^2 <
B_1$; parametric scattering is schematically drawn in
Fig.~\ref{fig:steadystate}(d).

At $\kk_p = 0$ the scattering thresholds read
\begin{equation}
  \label{eq:sc-thr}
  P_{1,2}(\kk) = \frac23 [\omega_p - \omega(\kk)] \mp
  \frac13 \sqrt{[\omega_p - \omega(\kk)]^2 - 3 [\gamma(\kk)]^2};
\end{equation}
these are the two nearly parabolic curves seen in
Fig.~\ref{fig:steadystate}(b).  For simplicity, let $\gamma(\kk)$ be
constant.  The scattering from $\kk_p = 0$ to $\kk_s \neq \kk_p$ is
impossible at $\omega_p - \omega(\kk_p) < \sqrt{3} \gamma$, i.\,e.\ in
the absence of bistability, because $\omega(\kk_s) > \omega(\kk_p)$.
On the other hand, $d P_1(\kk) / d [\omega_p - \omega(\kk)] > 0$ at
$\omega_p - \omega(\kk) > 2 \gamma$ and, hence, $P_1(\kk_s \, {\neq}
\, \kk_p) < B_1$, i.\,e.\ the scattering threshold is at the lower
branch of one-mode solutions.  Then, with due regard for all
scattering directions,
\begin{equation}
  \label{eq:sc-thr-min}
  P = \min_\kk P_1(\kk) = \gamma
  \quad \text{if} \quad
  \omega_p - \omega(\kk_p \, {=} \, 0) > 2 \gamma;
\end{equation}
the minimum is reached at $\kk = \kk_s$ where $\omega_p -
\omega(\kk_s) = 2 \gamma$.  Finally, one can make certain that
\begin{equation}
  \label{eq:sc-thr-bs}
  f(B_2) < f(P) < f(B_1)
\end{equation}
(see.~Fig.~\ref{fig:steadystate}), i.\,e.\ $P$ is always within the
bistable area if $\kk_p = 0$ and $\omega_p - \omega(\kk_p) > 2 \gamma$
and $\gamma$ is independent of $\kk$.

In the limit of small $\gamma$ and large $D = \omega_p -
\omega(\kk_p)$, threshold $P$ can be arbitrarily smaller than $B_1
\gtrsim D/3$.  It is guaranteed near $\kk_p = 0$, but, generally
speaking, can also be valid in a wide region of $\kk_p$ and $\omega_p$
where resonant scattering $(\kk_p, \kk_p) \to (\kk, 2 \kk_p - \kk)$
along the dispersion curve is
permitted.\cite{Whittaker05,Wouters07-thr} This brings up the question
on the evolution scenario at $f \gtrsim f(P)$.  Since the signal gain
rate at the threshold is zero [$\max_\kk \Gamma(\kk, P) = 0$], one can
presume a ``soft'' onset of the scattering in such a way that
steady-state amplitudes $|\psi(\kk)|$ do not exhibit discontinuity
near the threshold.  This scenario is analogous to the second-order
phase transition and, as such, is often opposed to a ``rigid''
development of instability at $f = f(B_1)$ that involves jump in
$|\psi(\kk_p)|$.\cite{Gippius04-epl, Gippius04-cm, Whittaker05,
  Gavrilov07-en, Wouters07-thr} In following Sec.~\ref{sec:catastrophe}
we prove that, although the increase of scattering ``signals'' can be
arbitrarily slow and smooth in its beginning, even at constant $f$ it
is followed by a blowup, and finally the blueshift compensates the
initial mismatch~$D$.

\section{Inevitability of the catastrophe}
\label{sec:catastrophe}

\subsection{Seeking for steady states}
\label{sec:seeking-steady-states}

Above we have shown that on reaching the threshold the pumped mode
starts to break up, giving rise to the growth of scattering
``signals''.  Now we seek for the conditions under which this process
stops, resulting in a new steady-state field distribution.

Taking Eq.~(\ref{eq:gp}) at $\kk = \kk_p$, one finds that $\psi_p
\equiv \psi(\kk_p)$ obeys the equation
\begin{multline}
  \label{eq:gp-kp}
  i \frac{d}{dt} \psi_p(t) = 
  [\omega(\kk_p) - i \gamma(\kk_p)]\psi_p(t)
  + f(t) e^{-i \omega_p t} \\
  {} + V |\psi_p(t)|^2 \psi_p(t)
  + 2V \psi_p(t) \sum_{\kk \neq \kk_p} \psi^* (\kk, t) \psi (\kk, t)
  \\ {} + V \psi_p^*(t) \sum_{\kk \neq \kk_p}
  \psi (2 \kk_p \, {-} \, \kk, t) \psi (\kk, t)
  \\ {} + V \sum_{\substack{
      \mathbf{q}_1 \neq \kk_p, ~ \mathbf{q}_2 \neq \kk_p \\ 
      \mathbf{q}_1 + \mathbf{q}_2 \neq 2\kk_p}}
  \psi^* (\mathbf{q}_1 \, {+} \, \mathbf{q}_2 \, {-} \, \kk_p, t) \,
  \psi(\mathbf{q}_1, t) \, \psi(\mathbf{q}_2, t).
\end{multline}
It is appropriate to neglect the last sum (that is independent of
$\psi_p$) on the assumption of a continuous $k$-space distribution of
scattered modes so that $|\psi(\kk)/\psi_p| \ll 1$ for each $\kk \neq
\kk_p$; note as well that related scattering processes can hardly obey
the conservation laws~(\ref{eq:cons2}).  The remaining terms take
account of all processes of the types $(\kk_p, \kk) \leftrightarrow
(\kk, \kk_p)$ and $(\kk_p, \kk_p) \leftrightarrow (\kk, 2 \kk_p -
\kk)$.

Let
\begin{equation}
  \label{eq:steady-def}
  \psi_p(t) = \bar\psi_p e^{i \phi_0} e^{-i \omega_p t},
  \quad
  \psi(\kk, t) = \bar\psi(\kk) e^{i \phi(\kk)}
  e^{-i \tilde\omega(\kk) t},
  \quad 
\end{equation}
where all $\bar \psi$ take real and nonnegative values.  Together with
Eq.~(\ref{eq:cons2}) it enables us to exclude high-frequency
oscillations ($\propto e^{-i \omega_p t}$) and transition effects and
write the equation for $\bar\psi_0$ in the form of
Eq.~(\ref{eq:one-mode}), i.\,e.\
\begin{equation}
  \label{eq:kp}
  \bar\psi_p^2 = \frac{f^2}{
    [\omega_p - \bar\omega - V \bar\psi_p^2 ]{\vphantom\mathstrut}^2 
    + \bar\gamma {\vphantom\mathstrut}^2},
\end{equation}
where
\begin{gather}
  \bar\omega = \omega(\kk_p) + 2V \sum_{\kk \neq \kk_p}
  \bar\psi^2(\kk) \hspace{3cm} \notag \\
  \label{eq:omega}
  \hspace{2cm} {} + V \sum_{\kk \neq \kk_p} \bar\psi(\kk)
  \bar\psi(2\kk_p - \kk) \cos \chi(\kk), \\
  \label{eq:gamma}
  \bar\gamma = \gamma(\kk_p) + V \sum_{\kk \neq \kk_p}
  \bar\psi(\kk) \bar\psi(2\kk_p - \kk) \sin \chi(\kk), \\
  \label{eq:chi}
  \chi(\kk) = \phi(\kk) + \phi(2\kk_p - \kk) - 2 \phi_0.
\end{gather}

Mean amplitude $\bar\psi_p$ depends on both the external field and the
distribution of amplitudes $\bar\psi(\kk)$ and phases $\chi(\kk)$ of
scattered states.  Values $\bar\omega$ and $\bar\gamma$ have the
meaning of the frequency and decay rate of the driven mode.  In
particular, $\bar\gamma - \gamma(\kk_p)$ represents its losses per
unit time through the scattering from $\kk = \kk_p$ to $\kk \neq
\kk_p$. Thus, (i) $\bar\gamma - \gamma(\kk_p) > 0$ as long as
$\sum_{\kk \neq \kk_p} \bar \psi(\kk) > 0$ and (ii) $\sin \chi(\kk) >
0$ as long as $\bar\psi(\kk) > 0$.

An equilibrium steady-state field distribution, if it exists, implies
conservation of energy under constant pumping; in particular, balance
of energy should be fulfilled between the driven mode and the set of
scattered modes.  It means that a virtual increase in $\bar \gamma$,
that is connected with a virtual growth of non-zero ``signals'',
should involve a decrease in the steady-state amplitude of the driven
mode that feeds them; hence,
\begin{equation}
  \label{eq:equilibrium}
  \frac{\delta \bar\psi_p^2} {\delta \bar\gamma} < 0.
\end{equation}
Indeed, the contrary would mean an increase in both the energy of
intra-cavity field (as the interaction is repulsive and
energy-conserving) and the energy lighting out of the cavity.

Now let us find the conditions under which an equilibrium can take
place in the sense of Eq.~(\ref{eq:equilibrium}).  Taking the
differential of Eq.~(\ref{eq:kp}), we have
\begin{equation}
  \label{eq:deltapsi}
  \delta \bar\psi_p^2 = 2 \bar\psi_p^2 \cdot \frac{\delta X}{Y},
\end{equation}
where
\begin{gather}
  \label{eq:X1}
  \delta X = (\omega_p - \bar\omega - V \bar\psi_p^2) \delta
  \bar\omega - \bar\gamma \delta \bar\gamma,\\
  \label{eq:Y}
  Y = (\omega_p - \bar\omega - V \bar\psi_p^2)^2
  - 2 V \bar\psi_p^2 (\omega_p - \bar\omega - V \bar\psi_p^2)
  + \bar\gamma^2.
\end{gather}
From Eq.~(\ref{eq:kp}) one sees that $\partial (f^2) / \partial
(\bar\psi_p^2) = Y$.  Accordingly, $Y$ turns to zero at
\begin{equation}
  \label{eq:bar-b}
  \bar\psi_p^2 = \bar B_{1,2} \equiv \frac23 (\omega_p - \bar\omega)
  \mp \frac13 \sqrt{(\omega_p - \bar\omega)^2 - 3\bar\gamma^2}.
\end{equation}
$\bar B_1$ is the catastrophe point in which the steady-state
amplitude exhibits discontinuity.  Let us find the minimum of $\delta
X / \delta \bar\gamma$ at $V \bar\psi_p^2 \leq \bar B_1$.  We have
\begin{equation}
  \label{eq:X3}
  \delta X \ge \frac{1}{3}
  \left(\omega_p - \bar\omega
    + \sqrt{(\omega_p - \bar\omega)^2 - 3 \bar \gamma^2}
  \right) \delta\bar\omega - \bar\gamma\delta\bar\gamma.
\end{equation}
Then assume that $\omega_p - \bar\omega > \sqrt{3}\bar\gamma$;
consequently,
\begin{equation}
  \label{eq:X2}
  \frac{1}{\bar\gamma} \frac{\delta X}{\delta \bar\gamma} >
  \frac{1}{\sqrt{3}}
  \frac{\delta\bar\omega}{\vphantom{\sqrt{3}} \delta\bar\gamma}
  - 1.
\end{equation}
Estimate $\delta \bar\omega / \delta \bar\gamma$ at $\delta \bar\gamma
> 0$, $\delta\chi(\kk) = 0$, and $\delta \bar\psi(\kk) \geq 0$ for
each $\kk \neq \kk_p$.  It is appropriate to assume that
\begin{equation}
  \label{eq:si-ratio}
  \sum_{\kk \neq \kk_p}
  \bar\psi(2\kk_p - \kk) \delta \bar\psi(\kk) 
  \leq
  \sum_{\kk \neq \kk_p} \bar\psi(\kk) \delta \bar\psi(\kk);
\end{equation}
equality means equal gain rates of ``signal'' and ``idler''.  Then we
have
\begin{equation}
  \label{eq:mean}
  \frac{\delta \bar\omega}{\delta \bar\gamma} \geq
  \frac{\sum_{\kk \neq \kk_p} u(\kk) \delta v(\kk)}
  {\sum_{\kk \neq \kk_p} \delta v(\kk)},
\end{equation}
where
\begin{equation}
  \label{eq:mean-1}
  u(\kk) = \frac{2 + \cos \chi(\kk)}{\sin \chi(\kk)}, \quad
  \delta v(\kk) = \sin
  \chi(\kk) \bar\psi(2\kk_p - \kk) \delta \bar\psi(\kk).
\end{equation}
Note, $\bar\psi(\kk) > 0$ implies $\sin \chi(\kk) > 0$.  Hence,
(\ref{eq:mean}) is the average of $u(\kk) > 0$ over the region with
$\delta v(\kk) > 0$ and, therefore,
\begin{equation}
  \label{eq:omega-delta}
  \frac{\delta \bar\omega}{\delta \bar\gamma} \geq 
  \min_\chi \left| \frac{2 + \cos \chi}{\sin \chi} \right|
  = \sqrt{3};
\end{equation}
the minimum is reached at $\chi = 2\pi / 3$.  Combining
Eqs.~(\ref{eq:X2}) and (\ref{eq:omega-delta}), we finally have
\begin{equation}
  \label{eq:nonstop}
  \frac{\delta \bar\psi_p^2}{\delta \bar\gamma} > 0 \quad
  \text{everywhere at} \quad P < V \bar\psi_p^2 <
  \bar B_1.
\end{equation}
The result obtained proves that (i) slightly above the threshold the
scattering involves growth of $\bar\psi_p$; and that (ii) balance of
energy between the pumped mode and scattered modes cannot be
established below the catastrophe point.

\subsection{Stability analysis}
\label{sec:stability-analysis}

Let us now analyze asymptotic stability of stationary
solutions~(\ref{eq:kp}).  Let
\begin{equation}
  \label{eq:fluct-def}
  \psi_p(t) = \bar\psi_p e^{-i \omega_p t + i \phi_0} + \psi_1
  e^{-i \Omega t} + \psi_2 e^{-i (2 \omega_p - \Omega) t}.
\end{equation}
Following the standard procedure, we substitute it into
Eq.~(\ref{eq:gp-kp}), take account of the steady-state equation for
$\bar\psi_p$ (so that corresponding terms disappear), and keep only
the first powers of $\psi_{1,2}$.  Equation of the type $Q e^{-i\Omega
  t} + R e^{-i(2\omega_p-\Omega)t} = 0$ yields two separate equations
$Q = 0$ and $R = 0$.  Then we have
\begin{equation}
  \label{eq:fluct}
  \begin{pmatrix}
    A - \Omega & C \\
    -C^* & 2\omega_p - A^* - \Omega
  \end{pmatrix}
  \begin{pmatrix}
    \psi_1 \\ \psi_2^*
  \end{pmatrix}
  = 0,
\end{equation}
where
\begin{gather}
  \label{eq:symm}
  A = \omega_0 - i\gamma_0 + 2V \left( \bar\psi_p^2 + S_1 \right), \\
  \label{eq:antisymm}
  C = V \left( \bar\psi_p^2 + \langle e^{i\chi} \rangle S_2 \right)
  e^{2i\phi_0},
\end{gather}
where, in turn, $\omega_0 = \omega(\kk_p)$ and $\gamma_0 =
\gamma(\kk_p)$ are the ``unperturbed'' quantities and
\begin{gather}
  \label{eq:steady-s}
  S_1 = \sum_{\kk \neq \kk_p} \bar\psi^2(\kk),
  \quad
  S_2 = \sum_{\kk \neq \kk_p} \bar\psi(\kk) \bar\psi(2\kk_p - \kk), \\
  \label{eq:steady-chi}
  \langle e^{i\chi} \rangle = \frac{1}{S_2}
  \sum_{\kk \neq \kk_p}
  \bar\psi(\kk) \bar\psi(2\kk_p - \kk) e^{i\chi(\kk)}.
\end{gather}
We solve the characteristic equation for $\Omega$ and then solve
equation $\Im \Omega = 0$ for $V\bar\psi_p^2$. The roots read
\begin{multline}
  \label{eq:thr-2}
  V\bar\psi_p^2 = \bar B'_{1,2} \equiv
  \frac23 \left( \omega_p - \bar\omega \right) 
  + S_2 \langle \cos \chi \rangle \\
  {} \mp \frac13 \sqrt{
    [\omega_p - \bar\omega + 3 S_2 \langle \cos \chi
    \rangle]^2_{\vphantom 2}
    + 3 S_2^2 [ 1 - \langle \cos \chi \rangle^2_{\vphantom 2}]
    - 3 \gamma_0^2};
\end{multline}
interval $\bar B'_1 < V\bar\psi_p < \bar B'_2$ is the forbidden zone
where $\Re\Omega = \omega_p$ and $\Im\Omega > 0$.  In the above
formula we have expressed $S_1$ in terms of $\omega_p - \bar\omega$,
$S_2$, and $\chi$ using Eq.~\eqref{eq:omega}.

Equation~\eqref{eq:thr-2} shows that despite a decrease in $\omega_p -
\bar\omega - \bar\gamma$ the system is still bistable.  It is not
allowed to accumulate large energy in scattered modes while keeping
$\bar\psi_p$ relatively small.  Normally, $\bar B'_1 < \bar
B_1^{\vphantom\prime}$ unless $S_2 \to 0$ and, hence, $\bar\gamma \to
\gamma_0$.

Let us now estimate the total blueshift $\bar\omega + V\bar\psi_p^2 -
\omega_0$ above the instability area.  Consider the ``uncompensated''
fraction of the pump detuning, i.\,e.\
\begin{equation}
  \label{eq:uncompensated}
  n = \frac{\omega_p - \left(\bar\omega + \bar B'_2\right)}
  {\omega_p - \omega_0}
\end{equation}
at $\gamma_0 \to 0$.  According to Eq.~(\ref{eq:thr-2}), $n < 0$ at
$\langle \cos \chi \rangle > 0$, but otherwise $n$ can be positive.
It reaches its maximum $n_\mathrm{max} = 1/2$ at $S_1 = S_2 =
(\omega_p - \omega_0) / 4V$ and $\chi(\kk) \to \pi$ for each $\kk$,
which, however, is impossible because it implies $(\bar\gamma -
\gamma_0) / (\bar\omega - \omega_0) \to 0$.  A more realistic case of
$\chi = 2 \pi / 3$ gives $n_\mathrm{max} \approx 0.18$.  Thus, on the
upper branch the total blueshift is close to or even exceeds the
initial pump detuning.

We arrive at the conclusion that even at $\omega_p - \omega_0 \gg
\gamma_0$ an equilibrium cannot be reached until most of the pump
frequency mismatch is compensated by the blueshift due to the
increased field, irrespective of the system parameters.

\subsection{Dynamics of scattered modes}
\label{sec:dynamics-scattered-modes}

So far we have had no assumptions about $\bar\omega$ and $\bar\gamma$
those were just free parameters.  However, we need to make a
suggestion regarding the dependence of $\bar\gamma$ on $\bar\psi_p$ in
view of the full system \eqref{eq:gp} rather than steady-state
equation \eqref{eq:kp}.

Let us consider the case of yet-uncompensated pump detuning, i.\,e.\
$\omega_p - \bar\omega - V \bar\psi_p^2 \gg \gamma_0$.  The signal
modes are distributed continuously and, thus,
$\bar\psi(\kk)/\bar\psi_p \ll 1$ for each $\kk$ but $S_1$ can be
comparable to $\bar\psi_p^2$.  Therefore the Bogoliubov approximation
[(\ref{eq:bogoliubov1}), (\ref{eq:bogoliubov2})] is valid, allowing us
to find renormalized eigenfrequencies $\Omega$.  Let, for
definiteness, $\gamma(\kk) = \gamma_0$ irrespective of $\kk$ and
$\kk_p = 0$ so that $S_1 = S_2 = S$.  Then we have
\begin{multline}
  \label{eq:lin-scattered}
  \Omega(\kk) = \omega_p - i\gamma_0 \\ {} \pm
  \sqrt{[
      \omega_p - \omega(\kk)
      - 2V \left( S + \bar\psi_p^2 \right)
    ]^2_{\vphantom p}
    - V^2_{\vphantom p} \bar\psi_p^4}
\end{multline}
for $\kk \neq \kk_p$.  Consider maximum gain rate $\Gamma = \max_\kk
\Im \Omega(\kk)$.  It is easy to see that $\Gamma$ grows with
$V\bar\psi_p^2$ up to
\begin{multline}
  \label{eq:gain-growth}
  V\bar\psi_p^2 \to Z \equiv \frac23
  \left[ \omega_p - \omega(\kk_p) - 2S \right] \\
  {} = \frac23 \left( \omega_p - \bar\omega \right) +
  \frac23 S \langle \cos \chi \rangle > \bar B'_1.
\end{multline}
Note as well that the unstable area is widening in the $k$-space up to
$V\bar\psi_p^2 = \bar B_1'$ [similar to
Fig.~\ref{fig:steadystate}(b)].  Hence, the worst-case ``dynamical''
assumption is that $\bar\gamma$ grows with $\bar\psi_p$ in the range
of growing $\Gamma$, i.\,e.\
\begin{equation}
  \label{eq:dyn-growth-1}
  \frac{d \bar\gamma} {d(\bar\psi_p^2)} > 0
  \quad \text{at} \quad \bar P < V \bar\psi_p^2 \le Z,
  \quad Z > \bar B'_1.
\end{equation}
This allows us to bring together all results of this section.

\subsection{Hierarchy of instabilities and a route to the catastrophe}
\label{sec:hier-inst}

In the course of its evolution, the condensate passes through a chain
of critical transformations.

The combination of Eqs.~(\ref{eq:nonstop}) and (\ref{eq:dyn-growth-1})
means the positive feedback loop between $\bar\psi_p$ and $\bar\gamma$
which forms at $V\bar\psi_p^2 = P$.  Since then both the pumped mode
and scattering signals start to grow hyperbolically.  However, at that
early stage the pumped mode is still stable ``by itself'' in the sense
that its growth is determined by the growth of signals, and the latter
can be arbitrarily small in the vicinity of the threshold.  This is
the latency period whose duration is determined by $f - f(P)$.

Later, at $V\bar\psi_p^2 = \bar B'_1$ [Eq.~\eqref{eq:thr-2}] the system
enters the strong positive feedback regime in which the driven mode
could no longer keep its intensity fixed even at constant ``signals''.
However, their gain rate still increases well above $V\bar\psi_p^2 =
\bar B'_1$ for each $\langle e^{i\chi} \rangle$ according to
Eq.~\eqref{eq:gain-growth}.  Consequently, at this stage the system
cannot get stabilized at whatever phase and $k$-space distribution of
scattered modes, and there is no way back.

Finally, at $V\bar\psi_p^2 = \bar B_1$ [Eq.~(\ref{eq:bar-b})] the
feedback loop is short-circuited in the condensate mode, resulting in
the explosive growth of its amplitude.  Here the system drastically
alters its state with respect to the external field and afterwards
comes to the equilibrium.

\section{Main results and discussion}
\label{sec:main-results}

\subsection{Catastrophic behavior}
\label{sec:evolution-scenario}

The results of the previous section allow one to understand the
dynamics of the condensate above the scattering threshold.  Let us
repeat the main points more informally.

The considered phenomena take place under above-resonance
excitation. The break-up $(\kk_p, \kk_p) \to (\kk, 2\kk_p - \kk)$ that
begins at the threshold $V\bar\psi_p^2 = P$ necessarily involves
further growth of the pumped mode $\bar\psi_p$, even if the pump $f^2$
remains constant.  This is because approaching the resonance (due to
the blueshift towards the pump frequency) compensates all additional
losses brought on by strengthening the scattering from the pumped
mode.  As a result, its decay rate ($\bar\gamma$) and frequency
($\bar\omega$) increase continuously, and the distribution of
scattered modes changes with time.  It appears that modes with
increasingly greater gain rates $\Gamma$ successively get involved in
scattering [see Fig.~\ref{fig:steadystate}(b), (d)].  At the same
time, the increase in either or both of $\bar\gamma$ and $\bar\omega$
lowers $\bar B_1$, the critical amplitude at which $d\bar\psi_p/dt$
tends to infinity.  The process of a ``smooth'' growth of $\bar\psi_p$
cannot stop before meeting the catastrophe point $V\bar\psi_p^2 = \bar
B_1$.  Therein the field blows up.  Finally equilibrium is reached
when the initial mismatch $\omega_p - \omega(\kk_p)$ gets compensated
by the polariton blueshift.

\subsection{OPO solutions}
\label{sec:opo-solutions}

The final state of the system can be different depending on the pump
wave vector~$\kk_p$.  If $\kk_p = 0$, \emph{all} steady states are
one-mode because the inter-mode scattering is impossible at $V
|\psi(\kk_p \, {=} \, 0)|^2 \ge B_2$~[see. Eq.~(\ref{eq:sc-thr})].  By
contrast, if $\kk_p$ is near the inflection point of the polariton
dispersion curve, then the final state can have macro-occupied
``signal'' ($\kk_s \approx 0$) and ``idler'' ($\kk_i \approx 2\kk_p$)
modes~\cite{Butte03,Krizhanovskii08} whose average populations remain
unchanged at a constant pump power, which is often referred to as
polariton optical parametric oscillation
(OPO).\cite{Whittaker01,Whittaker05}

Regarding the OPO states, our findings shed light on one of their
experimentally evidenced peculiarities.  Namely, it is known that
``signal'' and ``idler'' appear at $\kk_s \approx 0$ and $\kk_i
\approx 2\kk_p$ irrespective of the pump detuning $D = \omega_p -
\omega(\kk_p)$,\cite{Butte03} whereas the conservation
laws~(\ref{eq:cons2}) imply a broad `8'-shaped distribution of
scattered modes as long as $V\bar\psi_p^2 \gtrsim P$ and $D >
0$.\cite{Gippius04-epl,Gavrilov07-en,Krizhanovskii08} Such behavior
can, however, easily be understood in view of the discussed scenario:
the detuning $D$ is compensated by the blueshift of the dispersion
curve [Sec.~\ref{sec:stability-analysis}], which yields $\kk_s \approx
0$ irrespective of $D$.  In this respect our work supplements and
explains the findings by D.\,Whittaker who showed that three-mode OPO
states become asymptotically stable not earlier than the excitation
density reaches a certain comparatively large
magnitude.\cite{Whittaker05} Note as well that the hypothesis for the
crucial role of the bistability in OPO formation was made already in
Refs.~\onlinecite{Gippius04-epl,Gippius04-cm}.  In the experimental
work~\onlinecite{Krizhanovskii08} employing cw pump conditions it was
shown that a broad `8'-shaped distribution of scattered modes, that is
observed in the linear regime, immediately turns into the three-mode
OPO state on reaching the parametric threshold; in other words,
reaching the threshold is accompanied by a jump in the polariton
energy.  At the same time, essentially collective phenomena were shown
to play a role in OPO formation dynamics.\cite{Demenev08} Finally, in
this work we have found that it is general that reaching the
parametric scattering threshold at $D \gg \gamma$ involves sharp and
very significant jump in energy, and that such effect is essentially
collective.

\subsection{Accumulation of energy}
\label{sec:accumulation-energy}

Under a continuous-wave excitation the threshold of the transition
between steady states coincides with the scattering threshold and,
hence, is comparable to~$\gamma$ [see~(\ref{eq:sc-thr-min})].  At
$\kk_p = 0$ the threshold pump density is
\begin{equation}
  \label{eq:f-thr}
  f_\mathrm{thr}^2 = f^2 (P) = 
  \frac{\gamma}{V} \left[ (D - \gamma)^2 + \gamma^2 \right],
\end{equation}
where $D = \omega_p - \omega(\kk_p)$; the existence of the high-energy
solution at the same $f$ is ensured by Eq.~(\ref{eq:sc-thr-bs}).  Then
integrally
\begin{alignat}{2}
  \label{eq:lower}
  &V |\psi(\kk_p)|^2 < \gamma \quad &&\text{if} ~ f <
  f_\mathrm{thr}, \\
  \label{eq:upper}
  &V |\psi(\kk_p)|^2 > D \quad &&\text{if} ~ f > f_\mathrm{thr}.
\end{alignat}
Note that Eq.~(\ref{eq:upper}) is also characteristic of a one-mode
bistable oscillator (like that considered in
Ref.~\onlinecite{Baas04-pra}), but in that case the threshold
$f_\mathrm{thr}^2 = f^2(B_1)$ does not depend on $\gamma$ at $\gamma /
D \to 0$ and grows as $D^3 / V$ with increasing pump frequency.  By
contrast, we have shown that in a system with a high density of states
the threshold can be infinitesimally small at $\gamma \to 0$; and,
assuming that the threshold is reached, the steady-state response is
strong [$V|\psi|^2 \gtrsim D$] in a very wide range of pump
frequencies up to $D = 3 V f^2 / 2\gamma^2$ [so that $f^2 \ge f^2
(B_2)$].  Thus, even under a weak pump the energy is gradually
accumulated in the signal modes and with time becomes sufficient for
the transition to the upper steady-state branch.  Near the threshold,
$d \Gamma / d f^2 \to V / D^2$ at $\gamma \to 0$ [see
Fig.~\ref{fig:steadystate}(c) and Eq.~(\ref{eq:f-thr})], so the time
of energy accumulation can be large yet it does not tend to infinity
at $\gamma \to 0$ but is determined by the values of $f -
f_\mathrm{thr}$ and $D$.  Thus, it turns out that at $f(P) \lesssim f
< f(B_1)$ the pump intensity determines the latency period of the
transition rather than the eventual field amplitude.

\section{Numerical example}
\label{sec:num}

Below is an example of evolution of a resonantly pumped polariton
system whose parameters correspond to Fig.~\ref{fig:steadystate} at
above-threshold pump intensity $f^2$ such that
\begin{equation}
\label{eq:pump-num}
f^2(P) : f^2 : f^2(B_1) \approx 1 : 1.1 : 3.
\end{equation}

Equation~(\ref{eq:gp}) is solved on a square grid $(k_x, k_y)$ of
dimension $81 \times 81$ and size $-1.5 \leq k_{x,y} \leq
1.5~\mu\mathrm m^{-1}$.  The pump wave vector $\kk_p = 0$.  To
simulate scattering near the threshold, the right side of
Eq.~(\ref{eq:gp}) is supplemented by a stochastic term $\xi(\kk, t)$
that has the properties of white noise:
\begin{gather}
  \label{eq:noise}
  \langle \xi(\kk, t) \rangle = 0, \\
  \langle \xi^*(\kk_1, t_1) \, \xi(\kk_2, t_2) \rangle = a
  \delta(\kk_1, \kk_2) \, \delta(t_1, t_2)
\end{gather}
Its intensity $a$ is sufficient for creating average background
population $V|\psi(\kk)|^2 \approx 10^{-9} P$ at $f = 0$ for each
$\kk$, while its phase $\arg \xi(\kk, t)$ takes random values changing
each 80~fs at each $\kk$.  Polariton lifetime $\tau = 1/\gamma \approx
16$~ps; the full time interval on which Eq.~(\ref{eq:gp}) is solved
lasts $1100$~ps.  The pump detuning and resonance width are $\hbar
[\omega_p - \omega(\kk_p)] = 0.5$~meV and $\hbar\gamma \approx
0.04$~meV, respectively.

\begin{figure}
  \centering
  \includegraphics[width=1\linewidth]{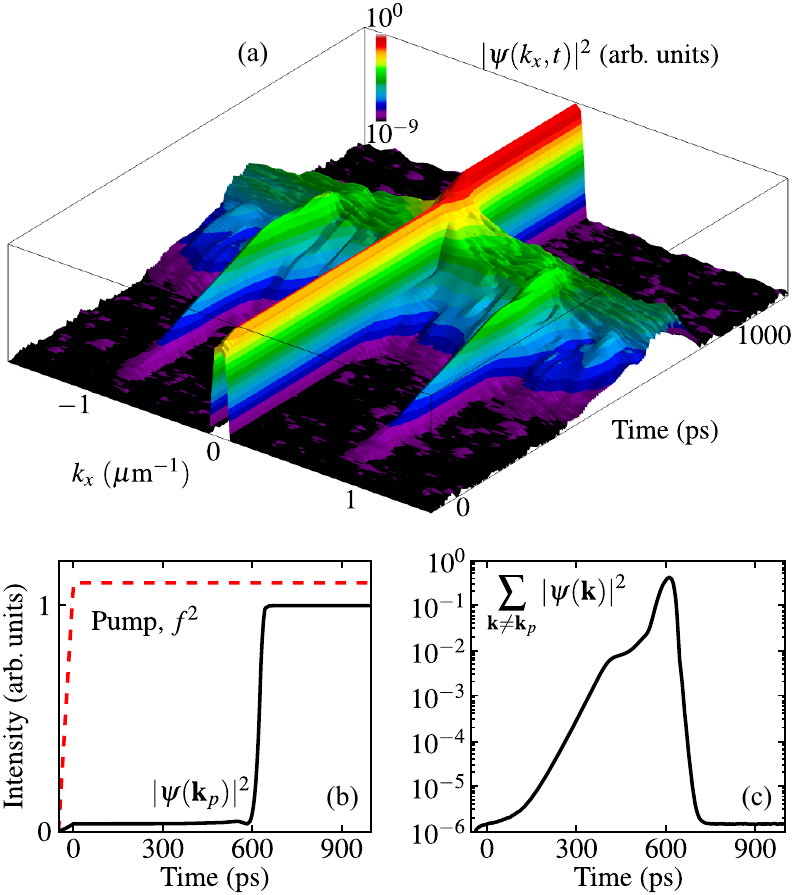}
  \caption{(a) Distribution of intra-cavity field $|\psi(k_x, t)|^2$
    at $k_y = 0$; (b) time dependences of $f^2$ (dashed line) and
    $|\psi(\kk_p)|^2$ (solid line); (c) time dependence of $\sum_{\kk
      \neq \kk_p} |\psi(\kk)|^2$.}
  \label{fig:kspace}
\end{figure}

Figure~\ref{fig:kspace}(a) presents the $k$-space and time
distribution of the field, $|\psi(k_x, t)|^2$ at $k_y = 0$.
Figure~\ref{fig:kspace}(b) shows the intensities of the pump $f^2$ and
the driven mode $|\psi(\kk_p)|^2$ as functions of time, and
Fig.~\ref{fig:kspace}(c) shows the integral intensity of scattered
modes.

Pump intensity $f^2$ increases linearly during 50~ps and then remains
constant starting from $t = 0$.  In the interval $0 < t \lesssim
100$~ps, steadily correlated ``signal'' modes appear out of the noise
substrate at $k_x^2 + k_y^2 = |\kk_s|^2$ [see also
Fig.~\ref{fig:steadystate}(b), (d)].  It is followed by the period of
their exponential growth at a constant rate $\Gamma \approx (1/3)
\gamma(\kk_s)$.  During this period ($100 \lesssim t \lesssim 400$~ps)
the signals are still weak and do not provide a significant feedback
to the pumped mode, yet by $t=400$~ps a slight increase in
$|\psi(\kk_p)|^2$ gets noticeable.

At $t \approx 400$~ps the system reaches threshold $V|\psi(\kk_p)|^2 =
\bar B_1'$ [Eq.~(\ref{eq:thr-2})].  The renormalized dispersion
surface now has a wide flat area $[\tilde\omega(\kk) = \omega_p]$
centered at $\kk = 0$.  Consequently, various scattering directions
become permitted by the conservation lows.  The scattering is
``disordered'', the old signals go away from resonance but the new
ones get out of the noise substrate and then grow quite rapidly.  At
$t=500$~ps, sum $\sum_{\kk \neq \kk_p} |\psi(\kk)|^2$ is only one
order of magnitude less than $|\psi(\kk_p)|^2$, the latter being 30\%
larger than in the beginning of the process at $t = 0$.

Finally, in the catastrophe point the field grows explosively, and the
system passes onto the upper steady-state branch ($t=600$~ps).
Henceforth the scattering to $\kk \neq \kk_p$ is no longer permitted
by the conservation laws, so all the scattered modes reduce down to
the noise level within the next 100~ps.  The new one-mode state is
stable and remains unchanged.

The scenario observed is general and reproduced irrespective of the
number of nodes and size of the computational grid which, however, may
strongly affect the characteristic evolution times.  An important
parameter is the noise amplitude $\xi$ that in real systems is
determined by quantum fluctuations and/or incoherent polariton states
filled due to Rayleigh scattering on inhomogeneities.  On one hand, $a
= \langle \xi^* \xi \rangle$ is the intensity level where the signal
rises out of the noise; thus, the larger $a$, the smaller the time
necessary for accumulating the critical energy in the course of a
forthcoming regular increase at a given rate $\Gamma$.  On the other
hand, in the vicinity of $f = f_\mathrm{thr}$ fluctuations may cause a
markable extension of a period when steadily correlated signal-idler
pairs get out of the noise substrate.  In our example it happens in
the interval $0 < t \lesssim 100$~ps and then at $400 \lesssim t
\lesssim 500$~ps.  To analyze fluctuations in a more tight connection
to their physical origins, one should either simulate inhomogeneities
and Rayleigh scattering\cite{Sekretenko13-fluct} or use a
probabilistic approach based on the Fokker-Planck
equations\cite{Maslova09,Johne09} or take account of the
exciton-phonon scattering and finite-temperature
effects.\cite{Bozat12}

\section{Concluding remarks}
\label{sec:conclusion}

In this work we have found that coherently driven Bose condensates
with a repulsive interaction can accumulate energy, which results in
blowup-like dynamical scenarios.  Above the threshold
(\ref{eq:f-thr}), the inter-particle interaction converts the
difference between the driving wave $(\omega_p)$ and condensate
$(\omega_0)$ frequency levels into the increased field $|\psi|^2 \sim
(\omega_p - \omega_0)/V$, so that the blue-shift of the resonance
compensates the mismatch of the pump frequency.

The discovered effect is expected to manifest itself in the
transitions between steady states in multistable cavity-polariton
systems.  It should become pronounced in high-$Q$ microcavities with
small decay rates $\gamma$.  The smaller $\gamma$, the wider the range
of pump frequencies $\omega_p$ in which intra-cavity field grows with
$\omega_p$ at a given pump power and, at the same time, the lower the
threshold density at a given $\omega_p$.  However, there is a tradeoff
between lowering the driving power and increasing the latency of the
transitions.  They still proceed at a comparatively weak pump but in
that case they appear to be delayed in time with respect to the moment
of reaching the threshold density [Fig.~\ref{fig:kspace}(b)].

Let us emphasize that the Gross-Pitaevskii equation is used on the
assumption of a macroscopically coherent state of the field.
Obviously, it cannot be true at whatever $\omega_p - \omega_0$ in real
finite-sized structures.  However, it is definitely valid in GaAs
based microcavities at $(\omega_p - \omega_0)/\gamma \gtrsim 10$ and
$Q \gtrsim 10^4$ and pump spot sizes of about $50~\mu$m, which follows
from a good agreement between modeled and measured
data.\cite{Gavrilov13-apl, Gavrilov13-mf, Sekretenko13-fluct} Note as
well the possibility for the interaction strength $V$ to depend on
$k_p$ and $\omega_p$.  In cavity-polariton systems, $V$ is determined
by the exciton Hopfield coefficient, so it is of no use to pump the
upper (photon-like) polariton branch far from resonance where $V$ is
nearly zero.  Next, the discussed scenario will no longer make sense
if the scattering threshold $P$ gets greater than $B_1$ [see
Fig.~\ref{fig:steadystate}(a)], which happens at large $k_p$ beyond
the ``magic angle'' [that is the inflection point in
$\omega_\mathrm{LP}(k)$, see Eq.~(\ref{eq:dispersion})].  Finally, in
this work we have neglected the spin degrees of freedom, assuming that
all polaritons have just the same spin.  Such assumption is
appropriate as long as opposite-spin polaritons interact comparatively
weakly, which, in its turn, normally appears to be the
case\cite{Renucci05} unless the pump frequency is very close to the
exciton level\cite{Vladimirova10,Sekretenko13-10ps} or TE/TM splitting
gets noticeable at large $k_p$ or a strong spatial anisotropy involves
splitting of the orthogonally polarized eigenstates even at $k =
0$.\cite{Gavrilov13-apl,Gavrilov13-mf,Sekretenko13-fluct}

The way of experimental verification of our findings is
straightforward under both cw and pulsed excitation conditions.  Under
a cw excitation, one can track the dependence of the threshold pump
density $(f^2_\mathrm{thr})$ on frequency detuning $D = \omega_p -
\omega(\kk_p)$ at $\kk_p = 0$. If our conclusions are correct,
$f^2_\mathrm{thr}$ at $D > 2\gamma$ should grow as $\gamma D^2$
[Eq.~(\ref{eq:f-thr})] instead of $D^3$; the cubic dependence was
predicted for the case of one-mode transitions.\cite{Baas04-pra} On
the other hand, the dynamics of the transitions between steady states
can be observed directly with the use of pulsed excitation and
time-resolved measurements on the scale of $10^2$~ps.  At
comparatively small pump densities the system should reveal the
latency period followed by a massive redistribution of scattered modes
in the far field.

\section*{Acknowledgements}

I wish to thank V.~D. Kulakovskii, N.~A. Gippius, and S.~G. Tikhodeev
for stimulating discussions.  2D and 3D visualizations are performed
using \textsc{MATPLOTLIB}\cite{Hunter07} and
\textsc{MAYAVI},\cite{Ramachandran11} respectively. I acknowledge
financial support by the RF President grant No.\ MK-6521.2014.2 and
the ``Dynasty'' foundation.

%


\begin{thebibliography}{42}%
\makeatletter
\providecommand \@ifxundefined [1]{%
 \@ifx{#1\undefined}
}%
\providecommand \@ifnum [1]{%
 \ifnum #1\expandafter \@firstoftwo
 \else \expandafter \@secondoftwo
 \fi
}%
\providecommand \@ifx [1]{%
 \ifx #1\expandafter \@firstoftwo
 \else \expandafter \@secondoftwo
 \fi
}%
\providecommand \natexlab [1]{#1}%
\providecommand \enquote  [1]{``#1''}%
\providecommand \bibnamefont  [1]{#1}%
\providecommand \bibfnamefont [1]{#1}%
\providecommand \citenamefont [1]{#1}%
\providecommand \href@noop [0]{\@secondoftwo}%
\providecommand \href [0]{\begingroup \@sanitize@url \@href}%
\providecommand \@href[1]{\@@startlink{#1}\@@href}%
\providecommand \@@href[1]{\endgroup#1\@@endlink}%
\providecommand \@sanitize@url [0]{\catcode `\\12\catcode `\$12\catcode
  `\&12\catcode `\#12\catcode `\^12\catcode `\_12\catcode `\%12\relax}%
\providecommand \@@startlink[1]{}%
\providecommand \@@endlink[0]{}%
\providecommand \url  [0]{\begingroup\@sanitize@url \@url }%
\providecommand \@url [1]{\endgroup\@href {#1}{\urlprefix }}%
\providecommand \urlprefix  [0]{URL }%
\providecommand \Eprint [0]{\href }%
\providecommand \doibase [0]{http://dx.doi.org/}%
\providecommand \selectlanguage [0]{\@gobble}%
\providecommand \bibinfo  [0]{\@secondoftwo}%
\providecommand \bibfield  [0]{\@secondoftwo}%
\providecommand \translation [1]{[#1]}%
\providecommand \BibitemOpen [0]{}%
\providecommand \bibitemStop [0]{}%
\providecommand \bibitemNoStop [0]{.\EOS\space}%
\providecommand \EOS [0]{\spacefactor3000\relax}%
\providecommand \BibitemShut  [1]{\csname bibitem#1\endcsname}%
\let\auto@bib@innerbib\@empty
\bibitem [{\citenamefont {Elesin}\ and\ \citenamefont
  {Kopaev}(1973)}]{Elesin73}%
  \BibitemOpen
  \bibfield  {author} {\bibinfo {author} {\bibfnamefont {V.~F.}\ \bibnamefont
  {Elesin}}\ and\ \bibinfo {author} {\bibfnamefont {Y.~V.}\ \bibnamefont
  {Kopaev}},\ }\href@noop {} {\bibfield  {journal} {\bibinfo  {journal} {Sov.
  Phys. JETP}\ }\textbf {\bibinfo {volume} {36}},\ \bibinfo {pages} {767}
  (\bibinfo {year} {1973})}\BibitemShut {NoStop}%
\bibitem [{\citenamefont {Baas}\ \emph {et~al.}(2004)\citenamefont {Baas},
  \citenamefont {Karr}, \citenamefont {Eleuch},\ and\ \citenamefont
  {Giacobino}}]{Baas04-pra}%
  \BibitemOpen
  \bibfield  {author} {\bibinfo {author} {\bibfnamefont {A.}~\bibnamefont
  {Baas}}, \bibinfo {author} {\bibfnamefont {J.~P.}\ \bibnamefont {Karr}},
  \bibinfo {author} {\bibfnamefont {H.}~\bibnamefont {Eleuch}}, \ and\ \bibinfo
  {author} {\bibfnamefont {E.}~\bibnamefont {Giacobino}},\ }\href {\doibase
  10.1103/PhysRevA.69.023809} {\bibfield  {journal} {\bibinfo  {journal} {Phys.
  Rev. A}\ }\textbf {\bibinfo {volume} {69}},\ \bibinfo {pages} {023809}
  (\bibinfo {year} {2004})}\BibitemShut {NoStop}%
\bibitem [{\citenamefont {Gippius}\ \emph {et~al.}(2004)\citenamefont
  {Gippius}, \citenamefont {Tikhodeev}, \citenamefont {Kulakovskii},
  \citenamefont {Krizhanovskii},\ and\ \citenamefont
  {Tartakovskii}}]{Gippius04-epl}%
  \BibitemOpen
  \bibfield  {author} {\bibinfo {author} {\bibfnamefont {N.~A.}\ \bibnamefont
  {Gippius}}, \bibinfo {author} {\bibfnamefont {S.~G.}\ \bibnamefont
  {Tikhodeev}}, \bibinfo {author} {\bibfnamefont {V.~D.}\ \bibnamefont
  {Kulakovskii}}, \bibinfo {author} {\bibfnamefont {D.~N.}\ \bibnamefont
  {Krizhanovskii}}, \ and\ \bibinfo {author} {\bibfnamefont {A.~I.}\
  \bibnamefont {Tartakovskii}},\ }\href
  {http://stacks.iop.org/0295-5075/67/i=6/a=997} {\bibfield  {journal}
  {\bibinfo  {journal} {EPL}\ }\textbf {\bibinfo {volume} {67}},\ \bibinfo
  {pages} {997} (\bibinfo {year} {2004})}\BibitemShut {NoStop}%
\bibitem [{\citenamefont {Gippius}\ and\ \citenamefont
  {Tikhodeev}(2004)}]{Gippius04-cm}%
  \BibitemOpen
  \bibfield  {author} {\bibinfo {author} {\bibfnamefont {N.~A.}\ \bibnamefont
  {Gippius}}\ and\ \bibinfo {author} {\bibfnamefont {S.~G.}\ \bibnamefont
  {Tikhodeev}},\ }\href {http://stacks.iop.org/0953-8984/16/i=35/a=007}
  {\bibfield  {journal} {\bibinfo  {journal} {Journal of Physics: Condensed
  Matter}\ }\textbf {\bibinfo {volume} {16}},\ \bibinfo {pages} {S3653}
  (\bibinfo {year} {2004})}\BibitemShut {NoStop}%
\bibitem [{\citenamefont {Para{\"\i}so}\ \emph {et~al.}(2010)\citenamefont
  {Para{\"\i}so}, \citenamefont {Wouters}, \citenamefont {L{\'e}ger},
  \citenamefont {Morier-Genoud},\ and\ \citenamefont
  {Deveaud-Pl{\'e}dran}}]{Paraiso10}%
  \BibitemOpen
  \bibfield  {author} {\bibinfo {author} {\bibfnamefont {T.~K.}\ \bibnamefont
  {Para{\"\i}so}}, \bibinfo {author} {\bibfnamefont {M.}~\bibnamefont
  {Wouters}}, \bibinfo {author} {\bibfnamefont {Y.}~\bibnamefont {L{\'e}ger}},
  \bibinfo {author} {\bibfnamefont {F.}~\bibnamefont {Morier-Genoud}}, \ and\
  \bibinfo {author} {\bibfnamefont {B.}~\bibnamefont {Deveaud-Pl{\'e}dran}},\
  }\href {\doibase 10.1038/nmat2787} {\bibfield  {journal} {\bibinfo  {journal}
  {Nat Mater}\ }\textbf {\bibinfo {volume} {9}},\ \bibinfo {pages} {655}
  (\bibinfo {year} {2010})}\BibitemShut {NoStop}%
\bibitem [{\citenamefont {Sarkar}\ \emph {et~al.}(2010)\citenamefont {Sarkar},
  \citenamefont {Gavrilov}, \citenamefont {Sich}, \citenamefont {Quilter},
  \citenamefont {Bradley}, \citenamefont {Gippius}, \citenamefont {Guda},
  \citenamefont {Kulakovskii}, \citenamefont {Skolnick},\ and\ \citenamefont
  {Krizhanovskii}}]{Sarkar10}%
  \BibitemOpen
  \bibfield  {author} {\bibinfo {author} {\bibfnamefont {D.}~\bibnamefont
  {Sarkar}}, \bibinfo {author} {\bibfnamefont {S.~S.}\ \bibnamefont
  {Gavrilov}}, \bibinfo {author} {\bibfnamefont {M.}~\bibnamefont {Sich}},
  \bibinfo {author} {\bibfnamefont {J.~H.}\ \bibnamefont {Quilter}}, \bibinfo
  {author} {\bibfnamefont {R.~A.}\ \bibnamefont {Bradley}}, \bibinfo {author}
  {\bibfnamefont {N.~A.}\ \bibnamefont {Gippius}}, \bibinfo {author}
  {\bibfnamefont {K.}~\bibnamefont {Guda}}, \bibinfo {author} {\bibfnamefont
  {V.~D.}\ \bibnamefont {Kulakovskii}}, \bibinfo {author} {\bibfnamefont
  {M.~S.}\ \bibnamefont {Skolnick}}, \ and\ \bibinfo {author} {\bibfnamefont
  {D.~N.}\ \bibnamefont {Krizhanovskii}},\ }\href {\doibase
  10.1103/PhysRevLett.105.216402} {\bibfield  {journal} {\bibinfo  {journal}
  {Phys. Rev. Lett.}\ }\textbf {\bibinfo {volume} {105}},\ \bibinfo {pages}
  {216402} (\bibinfo {year} {2010})}\BibitemShut {NoStop}%
\bibitem [{\citenamefont {Adrados}\ \emph {et~al.}(2010)\citenamefont
  {Adrados}, \citenamefont {Amo}, \citenamefont {Liew}, \citenamefont {Hivet},
  \citenamefont {Houdr{\'e}}, \citenamefont {Giacobino}, \citenamefont
  {Kavokin},\ and\ \citenamefont {Bramati}}]{Adrados10}%
  \BibitemOpen
  \bibfield  {author} {\bibinfo {author} {\bibfnamefont {C.}~\bibnamefont
  {Adrados}}, \bibinfo {author} {\bibfnamefont {A.}~\bibnamefont {Amo}},
  \bibinfo {author} {\bibfnamefont {T.~C.~H.}\ \bibnamefont {Liew}}, \bibinfo
  {author} {\bibfnamefont {R.}~\bibnamefont {Hivet}}, \bibinfo {author}
  {\bibfnamefont {R.}~\bibnamefont {Houdr{\'e}}}, \bibinfo {author}
  {\bibfnamefont {E.}~\bibnamefont {Giacobino}}, \bibinfo {author}
  {\bibfnamefont {A.~V.}\ \bibnamefont {Kavokin}}, \ and\ \bibinfo {author}
  {\bibfnamefont {A.}~\bibnamefont {Bramati}},\ }\href {\doibase
  10.1103/PhysRevLett.105.216403} {\bibfield  {journal} {\bibinfo  {journal}
  {Phys. Rev. Lett.}\ }\textbf {\bibinfo {volume} {105}},\ \bibinfo {pages}
  {216403} (\bibinfo {year} {2010})}\BibitemShut {NoStop}%
\bibitem [{\citenamefont {Gavrilov}\ \emph
  {et~al.}(2010{\natexlab{a}})\citenamefont {Gavrilov}, \citenamefont
  {Brichkin}, \citenamefont {Dorodnyi}, \citenamefont {Tikhodeev},
  \citenamefont {Gippius},\ and\ \citenamefont
  {Kulakovskii}}]{Gavrilov10-jetpl-en}%
  \BibitemOpen
  \bibfield  {author} {\bibinfo {author} {\bibfnamefont {S.}~\bibnamefont
  {Gavrilov}}, \bibinfo {author} {\bibfnamefont {A.}~\bibnamefont {Brichkin}},
  \bibinfo {author} {\bibfnamefont {A.}~\bibnamefont {Dorodnyi}}, \bibinfo
  {author} {\bibfnamefont {S.}~\bibnamefont {Tikhodeev}}, \bibinfo {author}
  {\bibfnamefont {N.}~\bibnamefont {Gippius}}, \ and\ \bibinfo {author}
  {\bibfnamefont {V.}~\bibnamefont {Kulakovskii}},\ }\href {\doibase
  10.1134/S0021364010150105} {\bibfield  {journal} {\bibinfo  {journal} {JETP
  Letters}\ }\textbf {\bibinfo {volume} {92}},\ \bibinfo {pages} {171}
  (\bibinfo {year} {2010}{\natexlab{a}})}\BibitemShut {NoStop}%
\bibitem [{\citenamefont {Gavrilov}\ \emph {et~al.}(2012)\citenamefont
  {Gavrilov}, \citenamefont {Brichkin}, \citenamefont {Demenev}, \citenamefont
  {Dorodnyy}, \citenamefont {Novikov}, \citenamefont {Kulakovskii},
  \citenamefont {Tikhodeev},\ and\ \citenamefont {Gippius}}]{Gavrilov12-prb}%
  \BibitemOpen
  \bibfield  {author} {\bibinfo {author} {\bibfnamefont {S.~S.}\ \bibnamefont
  {Gavrilov}}, \bibinfo {author} {\bibfnamefont {A.~S.}\ \bibnamefont
  {Brichkin}}, \bibinfo {author} {\bibfnamefont {A.~A.}\ \bibnamefont
  {Demenev}}, \bibinfo {author} {\bibfnamefont {A.~A.}\ \bibnamefont
  {Dorodnyy}}, \bibinfo {author} {\bibfnamefont {S.~I.}\ \bibnamefont
  {Novikov}}, \bibinfo {author} {\bibfnamefont {V.~D.}\ \bibnamefont
  {Kulakovskii}}, \bibinfo {author} {\bibfnamefont {S.~G.}\ \bibnamefont
  {Tikhodeev}}, \ and\ \bibinfo {author} {\bibfnamefont {N.~A.}\ \bibnamefont
  {Gippius}},\ }\href {\doibase 10.1103/PhysRevB.85.075319} {\bibfield
  {journal} {\bibinfo  {journal} {Phys. Rev. B}\ }\textbf {\bibinfo {volume}
  {85}},\ \bibinfo {pages} {075319} (\bibinfo {year} {2012})}\BibitemShut
  {NoStop}%
\bibitem [{\citenamefont {Weisbuch}\ \emph {et~al.}(1992)\citenamefont
  {Weisbuch}, \citenamefont {Nishioka}, \citenamefont {Ishikawa},\ and\
  \citenamefont {Arakawa}}]{Weisbuch92}%
  \BibitemOpen
  \bibfield  {author} {\bibinfo {author} {\bibfnamefont {C.}~\bibnamefont
  {Weisbuch}}, \bibinfo {author} {\bibfnamefont {M.}~\bibnamefont {Nishioka}},
  \bibinfo {author} {\bibfnamefont {A.}~\bibnamefont {Ishikawa}}, \ and\
  \bibinfo {author} {\bibfnamefont {Y.}~\bibnamefont {Arakawa}},\ }\href
  {\doibase 10.1103/PhysRevLett.69.3314} {\bibfield  {journal} {\bibinfo
  {journal} {Phys. Rev. Lett.}\ }\textbf {\bibinfo {volume} {69}},\ \bibinfo
  {pages} {3314} (\bibinfo {year} {1992})}\BibitemShut {NoStop}%
\bibitem [{\citenamefont {Yamamoto}\ \emph {et~al.}(2000)\citenamefont
  {Yamamoto}, \citenamefont {Tassone},\ and\ \citenamefont
  {Cao}}]{Yamamoto-book-2000}%
  \BibitemOpen
  \bibfield  {author} {\bibinfo {author} {\bibfnamefont {Y.}~\bibnamefont
  {Yamamoto}}, \bibinfo {author} {\bibfnamefont {T.}~\bibnamefont {Tassone}}, \
  and\ \bibinfo {author} {\bibfnamefont {H.}~\bibnamefont {Cao}},\ }\href@noop
  {} {\emph {\bibinfo {title} {{Semiconductor Cavity Quantum
  Electrodynamics}}}}\ (\bibinfo  {publisher} {Springer-Verlag},\ \bibinfo
  {year} {2000})\BibitemShut {NoStop}%
\bibitem [{\citenamefont {Kavokin}\ and\ \citenamefont
  {Malpuech}(2003)}]{Kavokin-book-03}%
  \BibitemOpen
  \bibfield  {author} {\bibinfo {author} {\bibfnamefont {A.~V.}\ \bibnamefont
  {Kavokin}}\ and\ \bibinfo {author} {\bibfnamefont {G.}~\bibnamefont
  {Malpuech}},\ }\href@noop {} {\emph {\bibinfo {title} {{Cavity
  Polaritons}}}}\ (\bibinfo  {publisher} {Elsevier},\ \bibinfo {address}
  {Amsterdam},\ \bibinfo {year} {2003})\BibitemShut {NoStop}%
\bibitem [{\citenamefont {Gippius}\ \emph {et~al.}(2007)\citenamefont
  {Gippius}, \citenamefont {Shelykh}, \citenamefont {Solnyshkov}, \citenamefont
  {Gavrilov}, \citenamefont {Rubo}, \citenamefont {Kavokin}, \citenamefont
  {Tikhodeev},\ and\ \citenamefont {Malpuech}}]{Gippius07}%
  \BibitemOpen
  \bibfield  {author} {\bibinfo {author} {\bibfnamefont {N.~A.}\ \bibnamefont
  {Gippius}}, \bibinfo {author} {\bibfnamefont {I.~A.}\ \bibnamefont
  {Shelykh}}, \bibinfo {author} {\bibfnamefont {D.~D.}\ \bibnamefont
  {Solnyshkov}}, \bibinfo {author} {\bibfnamefont {S.~S.}\ \bibnamefont
  {Gavrilov}}, \bibinfo {author} {\bibfnamefont {Y.~G.}\ \bibnamefont {Rubo}},
  \bibinfo {author} {\bibfnamefont {A.~V.}\ \bibnamefont {Kavokin}}, \bibinfo
  {author} {\bibfnamefont {S.~G.}\ \bibnamefont {Tikhodeev}}, \ and\ \bibinfo
  {author} {\bibfnamefont {G.}~\bibnamefont {Malpuech}},\ }\href {\doibase
  10.1103/PhysRevLett.98.236401} {\bibfield  {journal} {\bibinfo  {journal}
  {Phys. Rev. Lett.}\ }\textbf {\bibinfo {volume} {98}},\ \bibinfo {pages}
  {236401} (\bibinfo {year} {2007})}\BibitemShut {NoStop}%
\bibitem [{\citenamefont {Shelykh}\ \emph {et~al.}(2008)\citenamefont
  {Shelykh}, \citenamefont {Liew},\ and\ \citenamefont
  {Kavokin}}]{Shelykh08-prl}%
  \BibitemOpen
  \bibfield  {author} {\bibinfo {author} {\bibfnamefont {I.~A.}\ \bibnamefont
  {Shelykh}}, \bibinfo {author} {\bibfnamefont {T.~C.~H.}\ \bibnamefont
  {Liew}}, \ and\ \bibinfo {author} {\bibfnamefont {A.~V.}\ \bibnamefont
  {Kavokin}},\ }\href {\doibase 10.1103/PhysRevLett.100.116401} {\bibfield
  {journal} {\bibinfo  {journal} {Phys. Rev. Lett.}\ }\textbf {\bibinfo
  {volume} {100}},\ \bibinfo {pages} {116401} (\bibinfo {year}
  {2008})}\BibitemShut {NoStop}%
\bibitem [{\citenamefont {Gavrilov}\ \emph
  {et~al.}(2010{\natexlab{b}})\citenamefont {Gavrilov}, \citenamefont
  {Gippius}, \citenamefont {Tikhodeev},\ and\ \citenamefont
  {Kulakovskii}}]{Gavrilov10-en}%
  \BibitemOpen
  \bibfield  {author} {\bibinfo {author} {\bibfnamefont {S.~S.}\ \bibnamefont
  {Gavrilov}}, \bibinfo {author} {\bibfnamefont {N.~A.}\ \bibnamefont
  {Gippius}}, \bibinfo {author} {\bibfnamefont {S.~G.}\ \bibnamefont
  {Tikhodeev}}, \ and\ \bibinfo {author} {\bibfnamefont {V.~D.}\ \bibnamefont
  {Kulakovskii}},\ }\href {\doibase 10.1134/S1063776110050146} {\bibfield
  {journal} {\bibinfo  {journal} {JETP}\ }\textbf {\bibinfo {volume} {110}},\
  \bibinfo {pages} {825} (\bibinfo {year} {2010}{\natexlab{b}})}\BibitemShut
  {NoStop}%
\bibitem [{\citenamefont {Gavrilov}\ \emph
  {et~al.}(2013{\natexlab{a}})\citenamefont {Gavrilov}, \citenamefont
  {Sekretenko}, \citenamefont {Novikov}, \citenamefont {Schneider},
  \citenamefont {H{\"o}fling}, \citenamefont {Kamp}, \citenamefont {Forchel},\
  and\ \citenamefont {Kulakovskii}}]{Gavrilov13-apl}%
  \BibitemOpen
  \bibfield  {author} {\bibinfo {author} {\bibfnamefont {S.~S.}\ \bibnamefont
  {Gavrilov}}, \bibinfo {author} {\bibfnamefont {A.~V.}\ \bibnamefont
  {Sekretenko}}, \bibinfo {author} {\bibfnamefont {S.~I.}\ \bibnamefont
  {Novikov}}, \bibinfo {author} {\bibfnamefont {C.}~\bibnamefont {Schneider}},
  \bibinfo {author} {\bibfnamefont {S.}~\bibnamefont {H{\"o}fling}}, \bibinfo
  {author} {\bibfnamefont {M.}~\bibnamefont {Kamp}}, \bibinfo {author}
  {\bibfnamefont {A.}~\bibnamefont {Forchel}}, \ and\ \bibinfo {author}
  {\bibfnamefont {V.~D.}\ \bibnamefont {Kulakovskii}},\ }\href {\doibase
  10.1063/1.4773523} {\bibfield  {journal} {\bibinfo  {journal} {APL}\ }\textbf
  {\bibinfo {volume} {102}},\ \bibinfo {eid} {011104} (\bibinfo {year}
  {2013}{\natexlab{a}})}\BibitemShut {NoStop}%
\bibitem [{\citenamefont {Gavrilov}\ \emph
  {et~al.}(2013{\natexlab{b}})\citenamefont {Gavrilov}, \citenamefont
  {Sekretenko}, \citenamefont {Gippius}, \citenamefont {Schneider},
  \citenamefont {H{\"o}fling}, \citenamefont {Kamp}, \citenamefont {Forchel},\
  and\ \citenamefont {Kulakovskii}}]{Gavrilov13-mf}%
  \BibitemOpen
  \bibfield  {author} {\bibinfo {author} {\bibfnamefont {S.~S.}\ \bibnamefont
  {Gavrilov}}, \bibinfo {author} {\bibfnamefont {A.~V.}\ \bibnamefont
  {Sekretenko}}, \bibinfo {author} {\bibfnamefont {N.~A.}\ \bibnamefont
  {Gippius}}, \bibinfo {author} {\bibfnamefont {C.}~\bibnamefont {Schneider}},
  \bibinfo {author} {\bibfnamefont {S.}~\bibnamefont {H{\"o}fling}}, \bibinfo
  {author} {\bibfnamefont {M.}~\bibnamefont {Kamp}}, \bibinfo {author}
  {\bibfnamefont {A.}~\bibnamefont {Forchel}}, \ and\ \bibinfo {author}
  {\bibfnamefont {V.~D.}\ \bibnamefont {Kulakovskii}},\ }\href {\doibase
  10.1103/PhysRevB.87.201303} {\bibfield  {journal} {\bibinfo  {journal} {Phys.
  Rev. B}\ }\textbf {\bibinfo {volume} {87}},\ \bibinfo {pages} {201303}
  (\bibinfo {year} {2013}{\natexlab{b}})}\BibitemShut {NoStop}%
\bibitem [{\citenamefont {Sekretenko}\ \emph
  {et~al.}(2013{\natexlab{a}})\citenamefont {Sekretenko}, \citenamefont
  {Gavrilov}, \citenamefont {Novikov}, \citenamefont {Kulakovskii},
  \citenamefont {H{\"o}fling}, \citenamefont {Schneider}, \citenamefont
  {Kamp},\ and\ \citenamefont {Forchel}}]{Sekretenko13-fluct}%
  \BibitemOpen
  \bibfield  {author} {\bibinfo {author} {\bibfnamefont {A.~V.}\ \bibnamefont
  {Sekretenko}}, \bibinfo {author} {\bibfnamefont {S.~S.}\ \bibnamefont
  {Gavrilov}}, \bibinfo {author} {\bibfnamefont {S.~I.}\ \bibnamefont
  {Novikov}}, \bibinfo {author} {\bibfnamefont {V.~D.}\ \bibnamefont
  {Kulakovskii}}, \bibinfo {author} {\bibfnamefont {S.}~\bibnamefont
  {H{\"o}fling}}, \bibinfo {author} {\bibfnamefont {C.}~\bibnamefont
  {Schneider}}, \bibinfo {author} {\bibfnamefont {M.}~\bibnamefont {Kamp}}, \
  and\ \bibinfo {author} {\bibfnamefont {A.}~\bibnamefont {Forchel}},\ }\href
  {\doibase 10.1103/PhysRevB.88.205302} {\bibfield  {journal} {\bibinfo
  {journal} {Phys. Rev. B}\ }\textbf {\bibinfo {volume} {88}},\ \bibinfo
  {pages} {205302} (\bibinfo {year} {2013}{\natexlab{a}})}\BibitemShut
  {NoStop}%
\bibitem [{\citenamefont {Sekretenko}\ \emph
  {et~al.}(2013{\natexlab{b}})\citenamefont {Sekretenko}, \citenamefont
  {Gavrilov},\ and\ \citenamefont {Kulakovskii}}]{Sekretenko13-10ps}%
  \BibitemOpen
  \bibfield  {author} {\bibinfo {author} {\bibfnamefont {A.~V.}\ \bibnamefont
  {Sekretenko}}, \bibinfo {author} {\bibfnamefont {S.~S.}\ \bibnamefont
  {Gavrilov}}, \ and\ \bibinfo {author} {\bibfnamefont {V.~D.}\ \bibnamefont
  {Kulakovskii}},\ }\href {\doibase 10.1103/PhysRevB.88.195302} {\bibfield
  {journal} {\bibinfo  {journal} {Phys. Rev. B}\ }\textbf {\bibinfo {volume}
  {88}},\ \bibinfo {pages} {195302} (\bibinfo {year}
  {2013}{\natexlab{b}})}\BibitemShut {NoStop}%
\bibitem [{\citenamefont {Demenev}\ \emph {et~al.}(2008)\citenamefont
  {Demenev}, \citenamefont {Shchekin}, \citenamefont {Larionov}, \citenamefont
  {Gavrilov}, \citenamefont {Kulakovskii}, \citenamefont {Gippius},\ and\
  \citenamefont {Tikhodeev}}]{Demenev08}%
  \BibitemOpen
  \bibfield  {author} {\bibinfo {author} {\bibfnamefont {A.~A.}\ \bibnamefont
  {Demenev}}, \bibinfo {author} {\bibfnamefont {A.~A.}\ \bibnamefont
  {Shchekin}}, \bibinfo {author} {\bibfnamefont {A.~V.}\ \bibnamefont
  {Larionov}}, \bibinfo {author} {\bibfnamefont {S.~S.}\ \bibnamefont
  {Gavrilov}}, \bibinfo {author} {\bibfnamefont {V.~D.}\ \bibnamefont
  {Kulakovskii}}, \bibinfo {author} {\bibfnamefont {N.~A.}\ \bibnamefont
  {Gippius}}, \ and\ \bibinfo {author} {\bibfnamefont {S.~G.}\ \bibnamefont
  {Tikhodeev}},\ }\href {\doibase 10.1103/PhysRevLett.101.136401} {\bibfield
  {journal} {\bibinfo  {journal} {Phys. Rev. Lett.}\ }\textbf {\bibinfo
  {volume} {101}},\ \bibinfo {pages} {136401} (\bibinfo {year}
  {2008})}\BibitemShut {NoStop}%
\bibitem [{\citenamefont {Krizhanovskii}\ \emph {et~al.}(2008)\citenamefont
  {Krizhanovskii}, \citenamefont {Gavrilov}, \citenamefont {Love},
  \citenamefont {Sanvitto}, \citenamefont {Gippius}, \citenamefont {Tikhodeev},
  \citenamefont {Kulakovskii}, \citenamefont {Whittaker}, \citenamefont
  {Skolnick},\ and\ \citenamefont {Roberts}}]{Krizhanovskii08}%
  \BibitemOpen
  \bibfield  {author} {\bibinfo {author} {\bibfnamefont {D.~N.}\ \bibnamefont
  {Krizhanovskii}}, \bibinfo {author} {\bibfnamefont {S.~S.}\ \bibnamefont
  {Gavrilov}}, \bibinfo {author} {\bibfnamefont {A.~P.~D.}\ \bibnamefont
  {Love}}, \bibinfo {author} {\bibfnamefont {D.}~\bibnamefont {Sanvitto}},
  \bibinfo {author} {\bibfnamefont {N.~A.}\ \bibnamefont {Gippius}}, \bibinfo
  {author} {\bibfnamefont {S.~G.}\ \bibnamefont {Tikhodeev}}, \bibinfo {author}
  {\bibfnamefont {V.~D.}\ \bibnamefont {Kulakovskii}}, \bibinfo {author}
  {\bibfnamefont {D.~M.}\ \bibnamefont {Whittaker}}, \bibinfo {author}
  {\bibfnamefont {M.~S.}\ \bibnamefont {Skolnick}}, \ and\ \bibinfo {author}
  {\bibfnamefont {J.~S.}\ \bibnamefont {Roberts}},\ }\href {\doibase
  10.1103/PhysRevB.77.115336} {\bibfield  {journal} {\bibinfo  {journal} {Phys.
  Rev. B}\ }\textbf {\bibinfo {volume} {77}},\ \bibinfo {pages} {115336}
  (\bibinfo {year} {2008})}\BibitemShut {NoStop}%
\bibitem [{\citenamefont {Krizhanovskii}\ \emph {et~al.}(2013)\citenamefont
  {Krizhanovskii}, \citenamefont {Cerda-M{\'e}ndez}, \citenamefont {Gavrilov},
  \citenamefont {Sarkar}, \citenamefont {Guda}, \citenamefont {Bradley},
  \citenamefont {Santos}, \citenamefont {Hey}, \citenamefont {Biermann},
  \citenamefont {Sich}, \citenamefont {Fras},\ and\ \citenamefont
  {Skolnick}}]{Krizhanovskii13}%
  \BibitemOpen
  \bibfield  {author} {\bibinfo {author} {\bibfnamefont {D.~N.}\ \bibnamefont
  {Krizhanovskii}}, \bibinfo {author} {\bibfnamefont {E.~A.}\ \bibnamefont
  {Cerda-M{\'e}ndez}}, \bibinfo {author} {\bibfnamefont {S.~S.}\ \bibnamefont
  {Gavrilov}}, \bibinfo {author} {\bibfnamefont {D.}~\bibnamefont {Sarkar}},
  \bibinfo {author} {\bibfnamefont {K.}~\bibnamefont {Guda}}, \bibinfo {author}
  {\bibfnamefont {R.}~\bibnamefont {Bradley}}, \bibinfo {author} {\bibfnamefont
  {P.~V.}\ \bibnamefont {Santos}}, \bibinfo {author} {\bibfnamefont
  {R.}~\bibnamefont {Hey}}, \bibinfo {author} {\bibfnamefont {K.}~\bibnamefont
  {Biermann}}, \bibinfo {author} {\bibfnamefont {M.}~\bibnamefont {Sich}},
  \bibinfo {author} {\bibfnamefont {F.}~\bibnamefont {Fras}}, \ and\ \bibinfo
  {author} {\bibfnamefont {M.~S.}\ \bibnamefont {Skolnick}},\ }\href {\doibase
  10.1103/PhysRevB.87.155423} {\bibfield  {journal} {\bibinfo  {journal} {Phys.
  Rev. B}\ }\textbf {\bibinfo {volume} {87}},\ \bibinfo {pages} {155423}
  (\bibinfo {year} {2013})}\BibitemShut {NoStop}%
\bibitem [{\citenamefont {Egorov}\ \emph {et~al.}(2009)\citenamefont {Egorov},
  \citenamefont {Skryabin}, \citenamefont {Yulin},\ and\ \citenamefont
  {Lederer}}]{Egorov09}%
  \BibitemOpen
  \bibfield  {author} {\bibinfo {author} {\bibfnamefont {O.~A.}\ \bibnamefont
  {Egorov}}, \bibinfo {author} {\bibfnamefont {D.~V.}\ \bibnamefont
  {Skryabin}}, \bibinfo {author} {\bibfnamefont {A.~V.}\ \bibnamefont {Yulin}},
  \ and\ \bibinfo {author} {\bibfnamefont {F.}~\bibnamefont {Lederer}},\ }\href
  {\doibase 10.1103/PhysRevLett.102.153904} {\bibfield  {journal} {\bibinfo
  {journal} {Phys. Rev. Lett.}\ }\textbf {\bibinfo {volume} {102}},\ \bibinfo
  {pages} {153904} (\bibinfo {year} {2009})}\BibitemShut {NoStop}%
\bibitem [{\citenamefont {Egorov}\ \emph {et~al.}(2010)\citenamefont {Egorov},
  \citenamefont {Gorbach}, \citenamefont {Lederer},\ and\ \citenamefont
  {Skryabin}}]{Egorov10-prl}%
  \BibitemOpen
  \bibfield  {author} {\bibinfo {author} {\bibfnamefont {O.~A.}\ \bibnamefont
  {Egorov}}, \bibinfo {author} {\bibfnamefont {A.~V.}\ \bibnamefont {Gorbach}},
  \bibinfo {author} {\bibfnamefont {F.}~\bibnamefont {Lederer}}, \ and\
  \bibinfo {author} {\bibfnamefont {D.~V.}\ \bibnamefont {Skryabin}},\ }\href
  {\doibase 10.1103/PhysRevLett.105.073903} {\bibfield  {journal} {\bibinfo
  {journal} {Phys. Rev. Lett.}\ }\textbf {\bibinfo {volume} {105}},\ \bibinfo
  {pages} {073903} (\bibinfo {year} {2010})}\BibitemShut {NoStop}%
\bibitem [{\citenamefont {Sich}\ \emph {et~al.}(2014)\citenamefont {Sich},
  \citenamefont {Fras}, \citenamefont {Chana}, \citenamefont {Skolnick},
  \citenamefont {Krizhanovskii}, \citenamefont {Gorbach}, \citenamefont
  {Hartley}, \citenamefont {Skryabin}, \citenamefont {Gavrilov}, \citenamefont
  {Cerda-M{\'e}ndez}, \citenamefont {Biermann}, \citenamefont {Hey},\ and\
  \citenamefont {Santos}}]{Sich13}%
  \BibitemOpen
  \bibfield  {author} {\bibinfo {author} {\bibfnamefont {M.}~\bibnamefont
  {Sich}}, \bibinfo {author} {\bibfnamefont {F.}~\bibnamefont {Fras}}, \bibinfo
  {author} {\bibfnamefont {J.~K.}\ \bibnamefont {Chana}}, \bibinfo {author}
  {\bibfnamefont {M.~S.}\ \bibnamefont {Skolnick}}, \bibinfo {author}
  {\bibfnamefont {D.~N.}\ \bibnamefont {Krizhanovskii}}, \bibinfo {author}
  {\bibfnamefont {A.~V.}\ \bibnamefont {Gorbach}}, \bibinfo {author}
  {\bibfnamefont {R.}~\bibnamefont {Hartley}}, \bibinfo {author} {\bibfnamefont
  {D.~V.}\ \bibnamefont {Skryabin}}, \bibinfo {author} {\bibfnamefont {S.~S.}\
  \bibnamefont {Gavrilov}}, \bibinfo {author} {\bibfnamefont {E.~A.}\
  \bibnamefont {Cerda-M{\'e}ndez}}, \bibinfo {author} {\bibfnamefont
  {K.}~\bibnamefont {Biermann}}, \bibinfo {author} {\bibfnamefont
  {R.}~\bibnamefont {Hey}}, \ and\ \bibinfo {author} {\bibfnamefont {P.~V.}\
  \bibnamefont {Santos}},\ }\href {\doibase 10.1103/PhysRevLett.112.046403}
  {\bibfield  {journal} {\bibinfo  {journal} {Phys. Rev. Lett.}\ }\textbf
  {\bibinfo {volume} {112}},\ \bibinfo {pages} {046403} (\bibinfo {year}
  {2014})}\BibitemShut {NoStop}%
\bibitem [{\citenamefont {Stevenson}\ \emph {et~al.}(2000)\citenamefont
  {Stevenson}, \citenamefont {Astratov}, \citenamefont {Skolnick},
  \citenamefont {Whittaker}, \citenamefont {Emam-Ismail}, \citenamefont
  {Tartakovskii}, \citenamefont {Savvidis}, \citenamefont {Baumberg},\ and\
  \citenamefont {Roberts}}]{Stevenson00}%
  \BibitemOpen
  \bibfield  {author} {\bibinfo {author} {\bibfnamefont {R.~M.}\ \bibnamefont
  {Stevenson}}, \bibinfo {author} {\bibfnamefont {V.~N.}\ \bibnamefont
  {Astratov}}, \bibinfo {author} {\bibfnamefont {M.~S.}\ \bibnamefont
  {Skolnick}}, \bibinfo {author} {\bibfnamefont {D.~M.}\ \bibnamefont
  {Whittaker}}, \bibinfo {author} {\bibfnamefont {M.}~\bibnamefont
  {Emam-Ismail}}, \bibinfo {author} {\bibfnamefont {A.~I.}\ \bibnamefont
  {Tartakovskii}}, \bibinfo {author} {\bibfnamefont {P.~G.}\ \bibnamefont
  {Savvidis}}, \bibinfo {author} {\bibfnamefont {J.~J.}\ \bibnamefont
  {Baumberg}}, \ and\ \bibinfo {author} {\bibfnamefont {J.~S.}\ \bibnamefont
  {Roberts}},\ }\href {\doibase 10.1103/PhysRevLett.85.3680} {\bibfield
  {journal} {\bibinfo  {journal} {Phys. Rev. Lett.}\ }\textbf {\bibinfo
  {volume} {85}},\ \bibinfo {pages} {3680} (\bibinfo {year}
  {2000})}\BibitemShut {NoStop}%
\bibitem [{\citenamefont {Ciuti}\ \emph {et~al.}(2001)\citenamefont {Ciuti},
  \citenamefont {Schwendimann},\ and\ \citenamefont {Quattropani}}]{Ciuti01}%
  \BibitemOpen
  \bibfield  {author} {\bibinfo {author} {\bibfnamefont {C.}~\bibnamefont
  {Ciuti}}, \bibinfo {author} {\bibfnamefont {P.}~\bibnamefont {Schwendimann}},
  \ and\ \bibinfo {author} {\bibfnamefont {A.}~\bibnamefont {Quattropani}},\
  }\href {\doibase 10.1103/PhysRevB.63.041303} {\bibfield  {journal} {\bibinfo
  {journal} {Phys. Rev. B}\ }\textbf {\bibinfo {volume} {63}},\ \bibinfo
  {pages} {041303} (\bibinfo {year} {2001})}\BibitemShut {NoStop}%
\bibitem [{\citenamefont {Whittaker}(2001)}]{Whittaker01}%
  \BibitemOpen
  \bibfield  {author} {\bibinfo {author} {\bibfnamefont {D.~M.}\ \bibnamefont
  {Whittaker}},\ }\href {\doibase 10.1103/PhysRevB.63.193305} {\bibfield
  {journal} {\bibinfo  {journal} {Phys. Rev. B}\ }\textbf {\bibinfo {volume}
  {63}},\ \bibinfo {pages} {193305} (\bibinfo {year} {2001})}\BibitemShut
  {NoStop}%
\bibitem [{\citenamefont {Butt{\'e}}\ \emph {et~al.}(2003)\citenamefont
  {Butt{\'e}}, \citenamefont {Skolnick}, \citenamefont {Whittaker},
  \citenamefont {Bajoni},\ and\ \citenamefont {Roberts}}]{Butte03}%
  \BibitemOpen
  \bibfield  {author} {\bibinfo {author} {\bibfnamefont {R.}~\bibnamefont
  {Butt{\'e}}}, \bibinfo {author} {\bibfnamefont {M.~S.}\ \bibnamefont
  {Skolnick}}, \bibinfo {author} {\bibfnamefont {D.~M.}\ \bibnamefont
  {Whittaker}}, \bibinfo {author} {\bibfnamefont {D.}~\bibnamefont {Bajoni}}, \
  and\ \bibinfo {author} {\bibfnamefont {J.~S.}\ \bibnamefont {Roberts}},\
  }\href {\doibase 10.1103/PhysRevB.68.115325} {\bibfield  {journal} {\bibinfo
  {journal} {Phys. Rev. B}\ }\textbf {\bibinfo {volume} {68}},\ \bibinfo
  {pages} {115325} (\bibinfo {year} {2003})}\BibitemShut {NoStop}%
\bibitem [{\citenamefont {Whittaker}(2005)}]{Whittaker05}%
  \BibitemOpen
  \bibfield  {author} {\bibinfo {author} {\bibfnamefont {D.~M.}\ \bibnamefont
  {Whittaker}},\ }\href {\doibase 10.1103/PhysRevB.71.115301} {\bibfield
  {journal} {\bibinfo  {journal} {Phys. Rev. B}\ }\textbf {\bibinfo {volume}
  {71}},\ \bibinfo {pages} {115301} (\bibinfo {year} {2005})}\BibitemShut
  {NoStop}%
\bibitem [{\citenamefont {McLaughlin}\ \emph {et~al.}(1986)\citenamefont
  {McLaughlin}, \citenamefont {Papanicolaou}, \citenamefont {Sulem},\ and\
  \citenamefont {Sulem}}]{McLaughlin86}%
  \BibitemOpen
  \bibfield  {author} {\bibinfo {author} {\bibfnamefont {D.~W.}\ \bibnamefont
  {McLaughlin}}, \bibinfo {author} {\bibfnamefont {G.~C.}\ \bibnamefont
  {Papanicolaou}}, \bibinfo {author} {\bibfnamefont {C.}~\bibnamefont {Sulem}},
  \ and\ \bibinfo {author} {\bibfnamefont {P.~L.}\ \bibnamefont {Sulem}},\
  }\href {\doibase 10.1103/PhysRevA.34.1200} {\bibfield  {journal} {\bibinfo
  {journal} {Phys. Rev. A}\ }\textbf {\bibinfo {volume} {34}},\ \bibinfo
  {pages} {1200} (\bibinfo {year} {1986})}\BibitemShut {NoStop}%
\bibitem [{\citenamefont {Landman}\ \emph {et~al.}(1988)\citenamefont
  {Landman}, \citenamefont {Papanicolaou}, \citenamefont {Sulem},\ and\
  \citenamefont {Sulem}}]{Landman88}%
  \BibitemOpen
  \bibfield  {author} {\bibinfo {author} {\bibfnamefont {M.~J.}\ \bibnamefont
  {Landman}}, \bibinfo {author} {\bibfnamefont {G.~C.}\ \bibnamefont
  {Papanicolaou}}, \bibinfo {author} {\bibfnamefont {C.}~\bibnamefont {Sulem}},
  \ and\ \bibinfo {author} {\bibfnamefont {P.~L.}\ \bibnamefont {Sulem}},\
  }\href {\doibase 10.1103/PhysRevA.38.3837} {\bibfield  {journal} {\bibinfo
  {journal} {Phys. Rev. A}\ }\textbf {\bibinfo {volume} {38}},\ \bibinfo
  {pages} {3837} (\bibinfo {year} {1988})}\BibitemShut {NoStop}%
\bibitem [{\citenamefont {Carusotto}\ and\ \citenamefont
  {Ciuti}(2004)}]{Carusotto04}%
  \BibitemOpen
  \bibfield  {author} {\bibinfo {author} {\bibfnamefont {I.}~\bibnamefont
  {Carusotto}}\ and\ \bibinfo {author} {\bibfnamefont {C.}~\bibnamefont
  {Ciuti}},\ }\href {\doibase 10.1103/PhysRevLett.93.166401} {\bibfield
  {journal} {\bibinfo  {journal} {Phys. Rev. Lett.}\ }\textbf {\bibinfo
  {volume} {93}},\ \bibinfo {pages} {166401} (\bibinfo {year}
  {2004})}\BibitemShut {NoStop}%
\bibitem [{\citenamefont {Wouters}\ and\ \citenamefont
  {Carusotto}(2007)}]{Wouters07-thr}%
  \BibitemOpen
  \bibfield  {author} {\bibinfo {author} {\bibfnamefont {M.}~\bibnamefont
  {Wouters}}\ and\ \bibinfo {author} {\bibfnamefont {I.}~\bibnamefont
  {Carusotto}},\ }\href {\doibase 10.1103/PhysRevB.75.075332} {\bibfield
  {journal} {\bibinfo  {journal} {Phys. Rev. B}\ }\textbf {\bibinfo {volume}
  {75}},\ \bibinfo {pages} {075332} (\bibinfo {year} {2007})}\BibitemShut
  {NoStop}%
\bibitem [{\citenamefont {Gavrilov}\ \emph {et~al.}(2007)\citenamefont
  {Gavrilov}, \citenamefont {Gippius}, \citenamefont {Kulakovskii},\ and\
  \citenamefont {Tikhodeev}}]{Gavrilov07-en}%
  \BibitemOpen
  \bibfield  {author} {\bibinfo {author} {\bibfnamefont {S.~S.}\ \bibnamefont
  {Gavrilov}}, \bibinfo {author} {\bibfnamefont {N.~A.}\ \bibnamefont
  {Gippius}}, \bibinfo {author} {\bibfnamefont {V.~D.}\ \bibnamefont
  {Kulakovskii}}, \ and\ \bibinfo {author} {\bibfnamefont {S.~G.}\ \bibnamefont
  {Tikhodeev}},\ }\href {\doibase 10.1134/S1063776107050056} {\bibfield
  {journal} {\bibinfo  {journal} {JETP}\ }\textbf {\bibinfo {volume} {104}},\
  \bibinfo {pages} {715} (\bibinfo {year} {2007})}\BibitemShut {NoStop}%
\bibitem [{\citenamefont {Maslova}\ \emph {et~al.}(2009)\citenamefont
  {Maslova}, \citenamefont {Johne},\ and\ \citenamefont {Gippius}}]{Maslova09}%
  \BibitemOpen
  \bibfield  {author} {\bibinfo {author} {\bibfnamefont {N.}~\bibnamefont
  {Maslova}}, \bibinfo {author} {\bibfnamefont {R.}~\bibnamefont {Johne}}, \
  and\ \bibinfo {author} {\bibfnamefont {N.}~\bibnamefont {Gippius}},\ }\href
  {\doibase 10.1134/S0021364009120054} {\bibfield  {journal} {\bibinfo
  {journal} {JETP Letters}\ }\textbf {\bibinfo {volume} {89}},\ \bibinfo
  {pages} {614} (\bibinfo {year} {2009})}\BibitemShut {NoStop}%
\bibitem [{\citenamefont {Johne}\ \emph {et~al.}(2009)\citenamefont {Johne},
  \citenamefont {Maslova},\ and\ \citenamefont {Gippius}}]{Johne09}%
  \BibitemOpen
  \bibfield  {author} {\bibinfo {author} {\bibfnamefont {R.}~\bibnamefont
  {Johne}}, \bibinfo {author} {\bibfnamefont {N.~S.}\ \bibnamefont {Maslova}},
  \ and\ \bibinfo {author} {\bibfnamefont {N.~A.}\ \bibnamefont {Gippius}},\
  }\href {\doibase 10.1016/j.ssc.2008.12.032} {\bibfield  {journal} {\bibinfo
  {journal} {Solid State Communications}\ }\textbf {\bibinfo {volume} {149}},\
  \bibinfo {pages} {496} (\bibinfo {year} {2009})}\BibitemShut {NoStop}%
\bibitem [{\citenamefont {Bozat}\ \emph {et~al.}(2012)\citenamefont {Bozat},
  \citenamefont {Savenko},\ and\ \citenamefont {Shelykh}}]{Bozat12}%
  \BibitemOpen
  \bibfield  {author} {\bibinfo {author} {\bibfnamefont {{\"O}.}~\bibnamefont
  {Bozat}}, \bibinfo {author} {\bibfnamefont {I.~G.}\ \bibnamefont {Savenko}},
  \ and\ \bibinfo {author} {\bibfnamefont {I.~A.}\ \bibnamefont {Shelykh}},\
  }\href {\doibase 10.1103/PhysRevB.86.035413} {\bibfield  {journal} {\bibinfo
  {journal} {Phys. Rev. B}\ }\textbf {\bibinfo {volume} {86}},\ \bibinfo
  {pages} {035413} (\bibinfo {year} {2012})}\BibitemShut {NoStop}%
\bibitem [{\citenamefont {Renucci}\ \emph {et~al.}(2005)\citenamefont
  {Renucci}, \citenamefont {Amand}, \citenamefont {Marie}, \citenamefont
  {Senellart}, \citenamefont {Bloch}, \citenamefont {Sermage},\ and\
  \citenamefont {Kavokin}}]{Renucci05}%
  \BibitemOpen
  \bibfield  {author} {\bibinfo {author} {\bibfnamefont {P.}~\bibnamefont
  {Renucci}}, \bibinfo {author} {\bibfnamefont {T.}~\bibnamefont {Amand}},
  \bibinfo {author} {\bibfnamefont {X.}~\bibnamefont {Marie}}, \bibinfo
  {author} {\bibfnamefont {P.}~\bibnamefont {Senellart}}, \bibinfo {author}
  {\bibfnamefont {J.}~\bibnamefont {Bloch}}, \bibinfo {author} {\bibfnamefont
  {B.}~\bibnamefont {Sermage}}, \ and\ \bibinfo {author} {\bibfnamefont
  {K.~V.}\ \bibnamefont {Kavokin}},\ }\href {\doibase
  10.1103/PhysRevB.72.075317} {\bibfield  {journal} {\bibinfo  {journal} {Phys.
  Rev. B}\ }\textbf {\bibinfo {volume} {72}},\ \bibinfo {pages} {075317}
  (\bibinfo {year} {2005})}\BibitemShut {NoStop}%
\bibitem [{\citenamefont {Vladimirova}\ \emph {et~al.}(2010)\citenamefont
  {Vladimirova}, \citenamefont {Cronenberger}, \citenamefont {Scalbert},
  \citenamefont {Kavokin}, \citenamefont {Miard}, \citenamefont
  {Lema\^\i{}tre}, \citenamefont {Bloch}, \citenamefont {Solnyshkov},
  \citenamefont {Malpuech},\ and\ \citenamefont {Kavokin}}]{Vladimirova10}%
  \BibitemOpen
  \bibfield  {author} {\bibinfo {author} {\bibfnamefont {M.}~\bibnamefont
  {Vladimirova}}, \bibinfo {author} {\bibfnamefont {S.}~\bibnamefont
  {Cronenberger}}, \bibinfo {author} {\bibfnamefont {D.}~\bibnamefont
  {Scalbert}}, \bibinfo {author} {\bibfnamefont {K.~V.}\ \bibnamefont
  {Kavokin}}, \bibinfo {author} {\bibfnamefont {A.}~\bibnamefont {Miard}},
  \bibinfo {author} {\bibfnamefont {A.}~\bibnamefont {Lema\^\i{}tre}}, \bibinfo
  {author} {\bibfnamefont {J.}~\bibnamefont {Bloch}}, \bibinfo {author}
  {\bibfnamefont {D.}~\bibnamefont {Solnyshkov}}, \bibinfo {author}
  {\bibfnamefont {G.}~\bibnamefont {Malpuech}}, \ and\ \bibinfo {author}
  {\bibfnamefont {A.~V.}\ \bibnamefont {Kavokin}},\ }\href {\doibase
  10.1103/PhysRevB.82.075301} {\bibfield  {journal} {\bibinfo  {journal} {Phys.
  Rev. B}\ }\textbf {\bibinfo {volume} {82}},\ \bibinfo {pages} {075301}
  (\bibinfo {year} {2010})}\BibitemShut {NoStop}%
\bibitem [{\citenamefont {Hunter}(2007)}]{Hunter07}%
  \BibitemOpen
  \bibfield  {author} {\bibinfo {author} {\bibfnamefont {J.~D.}\ \bibnamefont
  {Hunter}},\ }\href {\doibase 10.1109/MCSE.2007.55} {\bibfield  {journal}
  {\bibinfo  {journal} {Computing in Science and Engineering}\ }\textbf
  {\bibinfo {volume} {9}},\ \bibinfo {pages} {90} (\bibinfo {year}
  {2007})}\BibitemShut {NoStop}%
\bibitem [{\citenamefont {Ramachandran}\ and\ \citenamefont
  {Varoquaux}(2011)}]{Ramachandran11}%
  \BibitemOpen
  \bibfield  {author} {\bibinfo {author} {\bibfnamefont {P.}~\bibnamefont
  {Ramachandran}}\ and\ \bibinfo {author} {\bibfnamefont {G.}~\bibnamefont
  {Varoquaux}},\ }\href {\doibase 10.1109/MCSE.2011.35} {\bibfield  {journal}
  {\bibinfo  {journal} {Computing in Science \& Engineering}\ }\textbf
  {\bibinfo {volume} {13}},\ \bibinfo {pages} {40} (\bibinfo {year}
  {2011})}\BibitemShut {NoStop}%
\end{thebibliography}

\end{document}